\documentclass{emulateapj}	%EmulateApJchange

\tabletypesize{\scriptsize}

\newcommand{\ParticleSizeCare}{0.5}
\newcommand{\SValue}{-0.02}
\newcommand{\SValueMinusOneSigma}{0.16}
\newcommand{\SValuePlusOneSigma}{-0.16}
\newcommand{\SValueMinusTwoSigma}{0.37}
\newcommand{\SValuePlusTwoSigma}{-0.27}
\newcommand{\SValueMinusThreeSigma}{0.68}
\newcommand{\SValuePlusThreeSigma}{-0.38}
\newcommand{\RatioBOTHCFHTKepler}{1.02}
\newcommand{\RatioBOTHErrorCFHTKepler}{0.20}
\newcommand{\ValueZeroCFHTAlphaI}{1.09}
\newcommand{\ValueZeroErrorCFHTAlphaI}{0.32}
\newcommand{\RatioCFHTCorrelatedI}{1.05}
\newcommand{\RatioErrorCFHTCorrelatedI}{0.36}
\newcommand{\RatioCFHTCorrelatedII}{0.97}
\newcommand{\RatioErrorCFHTCorrelatedII}{0.24}
\newcommand{\ValueZeroCFHTAlphaII}{1.23}
\newcommand{\ValueZeroErrorCFHTAlphaII}{0.27}
\newcommand{\ValueZerocfhtkeplerCFHTOne}{1.107}
\newcommand{\ValueZeroMinuscfhtkeplerCFHTOne}{0.276}
\newcommand{\ValueZeroPluscfhtkeplerCFHTOne}{0.293}
\newcommand{\ValueZeroThreeSigmaPluscfhtkeplerCFHTOne}{1.865}
\newcommand{\ValueTwocfhtkeplerCFHTOne}{3.881}
\newcommand{\ValueTwoMinuscfhtkeplerCFHTOne}{0.005}
\newcommand{\ValueTwoPluscfhtkeplerCFHTOne}{0.004}
\newcommand{\ValueThreecfhtkeplerCFHTOne}{0.00100}
\newcommand{\ValueThreeMinuscfhtkeplerCFHTOne}{0.00059}
\newcommand{\ValueThreePluscfhtkeplerCFHTOne}{0.00053}
\newcommand{\ValueZerocfhtkeplerKeplerOne}{1.036}
\newcommand{\ValueZeroMinuscfhtkeplerKeplerOne}{0.111}
\newcommand{\ValueZeroPluscfhtkeplerKeplerOne}{0.103}
\newcommand{\ValueZeroThreeSigmaPluscfhtkeplerKeplerOne}{1.368}
\newcommand{\ValueTwocfhtkeplerKeplerOne}{3.877}
\newcommand{\ValueTwoMinuscfhtkeplerKeplerOne}{0.003}
\newcommand{\ValueTwoPluscfhtkeplerKeplerOne}{0.002}
\newcommand{\ValueThreecfhtkeplerKeplerOne}{-0.00030}
\newcommand{\ValueThreeMinuscfhtkeplerKeplerOne}{0.00014}
\newcommand{\ValueThreePluscfhtkeplerKeplerOne}{0.00014}
\newcommand{\ValueZerocfhtkeplerCFHTTwo}{1.214}
\newcommand{\ValueZeroMinuscfhtkeplerCFHTTwo}{0.241}
\newcommand{\ValueZeroPluscfhtkeplerCFHTTwo}{0.193}
\newcommand{\ValueZeroThreeSigmaPluscfhtkeplerCFHTTwo}{1.936}
\newcommand{\ValueTwocfhtkeplerCFHTTwo}{22.841}
\newcommand{\ValueTwoMinuscfhtkeplerCFHTTwo}{0.005}
\newcommand{\ValueTwoPluscfhtkeplerCFHTTwo}{0.004}
\newcommand{\ValueThreecfhtkeplerCFHTTwo}{0.00056}
\newcommand{\ValueThreeMinuscfhtkeplerCFHTTwo}{0.00042}
\newcommand{\ValueThreePluscfhtkeplerCFHTTwo}{0.00046}
\newcommand{\ValueZerocfhtkeplerKeplerTwo}{1.268}
\newcommand{\ValueZeroMinuscfhtkeplerKeplerTwo}{0.103}
\newcommand{\ValueZeroPluscfhtkeplerKeplerTwo}{0.103}
\newcommand{\ValueZeroThreeSigmaPluscfhtkeplerKeplerTwo}{1.585}
\newcommand{\ValueTwocfhtkeplerKeplerTwo}{22.830}
\newcommand{\ValueTwoMinuscfhtkeplerKeplerTwo}{0.001}
\newcommand{\ValueTwoPluscfhtkeplerKeplerTwo}{0.001}
\newcommand{\ValueThreecfhtkeplerKeplerTwo}{0.00027}
\newcommand{\ValueThreeMinuscfhtkeplerKeplerTwo}{0.00014}
\newcommand{\ValueThreePluscfhtkeplerKeplerTwo}{0.00015}
\newcommand{\ValueZerohstkeplerHST}{0.058}
\newcommand{\ValueZeroMinushstkeplerHST}{0.066}
\newcommand{\ValueZeroPlushstkeplerHST}{0.051}
\newcommand{\ValueZeroThreeSigmaPlushstkeplerHST}{0.354}
\newcommand{\ValueTwohstkeplerHST}{180.337}

\newcommand{\ValueThreehstkeplerHST}{0.00007}
\newcommand{\ValueThreeMinushstkeplerHST}{0.00024}
\newcommand{\ValueThreePlushstkeplerHST}{0.00022}
\newcommand{\ValueZerohstkeplerKepler}{0.149}
\newcommand{\ValueZeroMinushstkeplerKepler}{0.102}
\newcommand{\ValueZeroPlushstkeplerKepler}{0.090}
\newcommand{\ValueZeroThreeSigmaPlushstkeplerKepler}{0.474}
\newcommand{\ValueTwohstkeplerKepler}{180.337}

\newcommand{\ValueThreehstkeplerKepler}{0.00146}
\newcommand{\ValueThreeMinushstkeplerKepler}{0.00016}
\newcommand{\ValueThreePlushstkeplerKepler}{0.00016}
\newcommand{\Ratiohstkepler}{0.39}
\newcommand{\RatioErrorhstkepler}{0.46}
\newcommand{\ValueZerorefstarA}{0.116}
\newcommand{\ValueZeroMinusrefstarA}{0.097}
\newcommand{\ValueZeroPlusrefstarA}{0.084}
\newcommand{\ValueZeroThreeSigmaPlusrefstarA}{0.375}
\newcommand{\ValueTworefstarA}{179.969}

\newcommand{\ValueThreerefstarA}{-0.35279}
\newcommand{\ValueThreeMinusrefstarA}{0.00016}
\newcommand{\ValueThreePlusrefstarA}{0.00022}
\newcommand{\ValueZerorefstarB}{0.012}
\newcommand{\ValueZeroMinusrefstarB}{0.021}
\newcommand{\ValueZeroPlusrefstarB}{0.011}
\newcommand{\ValueZeroThreeSigmaPlusrefstarB}{0.112}
\newcommand{\ValueTworefstarB}{179.969}

\newcommand{\ValueThreerefstarB}{-0.89589}
\newcommand{\ValueThreeMinusrefstarB}{0.00010}
\newcommand{\ValueThreePlusrefstarB}{0.00010}
\newcommand{\ValueZerorefstarC}{0.052}
\newcommand{\ValueZeroMinusrefstarC}{0.053}
\newcommand{\ValueZeroPlusrefstarC}{0.030}
\newcommand{\ValueZeroThreeSigmaPlusrefstarC}{0.183}
\newcommand{\ValueTworefstarC}{179.969}

\newcommand{\ValueThreerefstarC}{-0.96580}
\newcommand{\ValueThreeMinusrefstarC}{0.00010}
\newcommand{\ValueThreePlusrefstarC}{0.00010}
\newcommand{\ValueZerorefstarD}{0.030}
\newcommand{\ValueZeroMinusrefstarD}{0.039}
\newcommand{\ValueZeroPlusrefstarD}{0.024}
\newcommand{\ValueZeroThreeSigmaPlusrefstarD}{0.158}
\newcommand{\ValueTworefstarD}{179.969}

\newcommand{\ValueThreerefstarD}{-0.97340}
\newcommand{\ValueThreeMinusrefstarD}{0.00011}
\newcommand{\ValueThreePlusrefstarD}{0.00011}
\newcommand{\ValueZerorefstarE}{0.065}
\newcommand{\ValueZeroMinusrefstarE}{0.056}
\newcommand{\ValueZeroPlusrefstarE}{0.043}
\newcommand{\ValueZeroThreeSigmaPlusrefstarE}{0.209}
\newcommand{\ValueTworefstarE}{179.969}

\newcommand{\ValueThreerefstarE}{-0.98720}
\newcommand{\ValueThreeMinusrefstarE}{0.00012}
\newcommand{\ValueThreePlusrefstarE}{0.00010}
\newcommand{\ValueZerorefstarF}{0.009}
\newcommand{\ValueZeroMinusrefstarF}{0.018}
\newcommand{\ValueZeroPlusrefstarF}{0.017}
\newcommand{\ValueZeroThreeSigmaPlusrefstarF}{0.117}
\newcommand{\ValueTworefstarF}{179.969}

\newcommand{\ValueThreerefstarF}{-0.98967}
\newcommand{\ValueThreeMinusrefstarF}{0.00010}
\newcommand{\ValueThreePlusrefstarF}{0.00010}
\shorttitle{Multiwavelength observations of KIC 12557548b} 
\shortauthors{Croll et al.}

\begin{document}

\title{Multiwavelength Observations of the Candidate Disintegrating sub-Mercury KIC 12557548b\altaffilmark{*} } % \altaffilmark{**} \altaffilmark{***}
\author{Bryce Croll\altaffilmark{1},\altaffilmark{2}
Saul Rappaport\altaffilmark{1},
John DeVore\altaffilmark{3},
Ronald L. Gilliland\altaffilmark{4},
Justin R. Crepp\altaffilmark{5},
Andrew W. Howard\altaffilmark{6},
Kimberly M. Star\altaffilmark{4},
Eugene Chiang\altaffilmark{7},
Alan M. Levine\altaffilmark{1},
% Joshua N. Winn\altaffilmark{1},
%Ray Jayawardhana\altaffilmark{8},
Jon M. Jenkins\altaffilmark{8},
Loic Albert\altaffilmark{9},
Aldo S. Bonomo\altaffilmark{10},
Jonathan J. Fortney\altaffilmark{11},
Howard Isaacson\altaffilmark{12}
}

%David Lafreniere\altaffilmark{10}
% Claire Moutou\altaffilmark{13}
% Magali Deleuil\altaffilmark{12},

\altaffiltext{1}{Kavli Institute for Astrophysics and Space Research, Massachusetts Institute
of Technology, Cambridge, MA 02139, USA; croll@space.mit.edu}

\altaffiltext{2}{NASA Sagan Fellow}

\altaffiltext{3}{Visidyne, Inc., Santa Barbara, CA 93105, USA}

\altaffiltext{4}{Center for Exoplanets and Habitable Worlds, The Pennsylvania State University, 525 Davey Lab, University Park, PA 16802, USA}

\altaffiltext{5}{University of Notre Dame, Department of Physics, 225 Nieuwland Science Hall, Notre Dame, IN 46556, USA}

\altaffiltext{6}{Institute for Astronomy, University of Hawaii at Manoa, 2680 Woodlawn Drive, Honolulu, Hawaii 96822, USA}

\altaffiltext{7}{Departments of Astronomy and of Earth and Planetary Science, University of California at Berkeley, Hearst Field Annex B-20, Berkeley, CA 94720-3411, USA}

% \altaffiltext{8}{Department of Astronomy and Astrophysics, University of Toronto, 50 St. George Street, Toronto, ON M5S 3H4, Canada}

\altaffiltext{8}{SETI Institute/NASA Ames Research Center, M/S 244-30, Moffett Field, CA 94035, USA}

\altaffiltext{9}{D\'epartement de physique, Universit\'e de Montr\'eal, C.P.
6128 Succ. Centre-Ville, Montr\'eal, QC, H3C 3J7, Canada}

\altaffiltext{10}{INAF - Osservatorio Astrofisico di Torino, via Osservatorio 20, 10025 Pino Torinese, Italy}

\altaffiltext{11}{Department of Astronomy and Astrophysics, University of California, Santa Cruz, CA, 95064}

\altaffiltext{12}{Department of Astronomy, University of California, Berkeley, CA 94720}

% \altaffiltext{13}{Canada-France-Hawaii Telescope Corporation, 65-1238 Mamalahoa Highway, Kamuela, HI 96743, USA}

% \altaffiltext{12}{Aix Marseille University, CNRS, LAM (Laboratoire d'Astrophysique de Marseille), UMR 7326, 13388 Marseille cedex 13, France}

\altaffiltext{*}{Based on observations obtained with WIRCam, a joint project of CFHT, Taiwan, Korea, Canada, France,
at the Canada-France-Hawaii Telescope (CFHT) which is operated by the National Research Council (NRC) of Canada, the
Institute National des Sciences de l'Univers of the Centre National de la Recherche Scientifique of France, and the
University of Hawaii.}

% \altaffiltext{**}{Based on observations made with the NASA/ESA {\it Hubble Space Telescope},
% obtained at the Space Telescope Science Institute,
% which is operated by the Association of Universities for Research in Astronomy, Inc., under NASA contract NAS 5-26555.
% These observations are associated with program \#GO-12987.}

% \altaffiltext{***}{
% Some of the data presented herein were obtained at the W.M. Keck Observatory, which is operated as a scientific partnership among
% the California Institute of Technology, the University of California and the National Aeronautics and Space Administration.
% The Observatory was made possible by the generous financial support of the W.M. Keck Foundation.}

\begin{abstract}

We present multiwavelength photometry, 
high angular resolution imaging,
and radial velocities, of the unique and confounding 
disintegrating low-mass planet candidate KIC 12557548b.
Our high angular resolution imaging, which includes spacebased {\it HST}/WFC3 observations in the optical
($\sim$0.53 $\mu m$ and $\sim$0.77 $\mu m$),
and groundbased Keck/NIRC2 observations in K'-band ($\sim$2.12 $\mu m$), allow us to rule-out
background and foreground candidates at angular separations greater than 0.2\arcsec \
that are bright enough to be responsible for the transits we associate with KIC 12557548.
Our radial velocity limit from Keck/HIRES allows us to rule-out bound, low-mass stellar companions ($\sim$0.2
M$_\odot$) to KIC 12557548 on orbits less than 10 years,
as well as placing an upper-limit on the mass of the candidate 
planet of 1.2 Jupiter masses;	%PLANETMASSLIMIT
therefore, the combination
of our radial velocities, high angular-resolution
imaging, and photometry are able to rule-out most false positive interpretations of the transits.
Our precise multiwavelength photometry includes
two simultaneous detections of the transit of KIC 12557548b using CFHT/WIRCam at 2.15 $\mu m$ and the {\it Kepler} space
telescope at 0.6 $\mu m$, as well as simultaneous null-detections of the transit by
{\it Kepler} and {\it HST}/WFC3 at 1.4 $\mu m$.
Our simultaneous {\it HST}/WFC3 and {\it Kepler} null-detections, provide
no evidence for radically different transit depths at these wavelengths.
Our simultaneous CFHT/WIRCam detections in the near-infrared and with {\it Kepler} in the optical 
reveal very similar transit depths (the average ratio of the transit depths at $\sim$2.15 $\mu m$ compared 
to $\sim$0.6 $\mu m$ is: \RatioBOTHCFHTKepler \ $\pm$ \RatioBOTHErrorCFHTKepler). This suggests that if the transits
we observe are due to scattering from single-size particles streaming from the planet in a comet-like tail,
then the particles must be $\sim$\ParticleSizeCare \ microns in radius or larger, which would favour that KIC 12557548b is
a sub-Mercury, 
rather than super-Mercury, mass planet.

\end{abstract}


%Need to do a plot that compares the depth with aperture radii

\section{Introduction}

\citet{Rappaport12} presented intriguing, perplexing and downright peculiar {\it Kepler} observations
of the K-dwarf star KIC 12557548.
The {\it Kepler} space telescope's \citep{Borucki09} observations of this star
displayed repeating dips
every $\sim$15.7 hours that varied in depth from a maximum of $\sim$1.3\% of the stellar flux
to a minimum of $\sim$0.2\% or less 
without a discernible rhyme or reason to explain the depth variations.
In addition, the occultations were not the iconic transit-like shape we have come to expect
from extrasolar planets or binary stars,
but exhibited an obvious ingress/egress asymmetry, with a sharp ingress followed by 
a longer, more gradual egress.
A non-detection of ellipsoidal light variations allowed an upper-limit on the mass of the 
occulting object to be set at 3 Jupiter masses, and thus 
prompted the question of what was causing this odd photometry?
%After ruling out alternative possibilities of a planet precessing due to an unseen companion
The answer the authors proposed was that the peculiar {\it Kepler} observations of KIC 12557548
(hereafter KIC 1255) are due to a gradually disintegrating
low mass (super-Mercury) planet, KIC 12557548b (hereafter KIC 1255b).
The thought process is that this putative planet, with its extremely short orbital period, is
being roasted by its host star,
and is throwing off material in
fits and starts; at each passage in front of its parent star, the 
different amount of material being discarded by the planet leads to differences in the
resulting optical depth, thus explaining
the obvious transit depth variations.
The clear ingress/egress asymmetry of the transit is then due to the fact that the 
material is streaming behind the planet, forming
a long comet-like tail that obscures the star for a larger fraction of the orbit.
%Something about the variable optical depth.

Naturally, observations as odd as those presented by \citet{Rappaport12} and an explanation as exotic as a
disintegrating super-Mercury, invited a great deal of skepticism from the astronomical community.
Alternative theories that have been
discussed to explain the observed photometry include: (i) a bizarre
{\it Kepler} photometric artifact; (ii) a background blended eclipsing 
binary\footnote{Although how this would explain the ingress/egress
asymmetry, or the transit depth variations remains a mystery.};
(iii) an exotically chaotic triple; or (iv) a binary that is orbiting KIC
12557548 wherein one member of the binary system is a white dwarf fed by
an accretion disk \citep{Rappaport12}.

Inspired by the \citet{Rappaport12} result there have been a number of modelling efforts to interpret
the bizarre {\it Kepler} observations that seem to 
reinforce the possibility that the photometry of KIC 1255 is caused
by scattering off material streaming from a disintegrating low-mass planet.
Dust scattering models confirm the viability of the 
disintegrating planet scenario featuring a comet-like tail trailing the planet, composed of
sub-micron sized grains \citep{Brogi12}, or up to one micron (0.1 - 1.0 $\mu m$) sized grains \citep{Budaj13}.
These efforts suggest that the minute brightening just prior
to transit % \footnote{We note we use the terms occultation and transit interchangeably in this draft.} 
can be readily explained by enhanced forward scattering
from this dust cloud, while the ingress/egress asymmetry can be explained by a comet-like dust tail that has a 
particle density or size distribution that decreases with distance from the planet.
The richness of the {\it Kepler} data on KIC 1255b have led to suggestions of 
evolution of the cometary tail \citep{Budaj13}, and that the comet is best explained
by a two component model, with a dense coma and inner-tail and a diffuse outer tail \citep{Budaj13,vanWerkhoven13}.
Another effort by \citet{Kawahara13} suggests
that the observed transit depth variability 
may correlate with the stellar rotation period, and thus
the presumed variable mass loss rate of the planet may be a byproduct of the stellar activity,
specifically ultraviolet and X-ray radiation.
\citet{PerezBeckerChiang13} argue from the results of a hydrodynamical wind model, that we may be observing  
the final death-throes of a planet catastrophically evaporating, and that KIC 1255b
may range in mass from 0.02 - 0.07 $M_\oplus$ (less than twice that of the Moon, to
greater than Mercury), although for most solutions the mass of KIC 1255b is less than that of Mercury.
We note that subsequent to the submission of this work, there has been an announcement of a second low-mass planet candidate,
possibly hosting a comet-like tail \citep{Rappaport14}.

%INSERTHERE SHOULD I MENTION THAT I AM WORKING ON THIS.

One proposed method for elucidating the unknown nature of the material that is supposedly occulting 
KIC 1255, is multiwavelength simultaneous observations of the transit of the object.
As the efficiency of scattering diminishes for wavelengths longer than the approximate particle size \citep{HansenTravis74},
and given the inference of sub-micron size grains in the dust tail of this object \citep{Brogi12,Budaj13}, one might expect
that infrared
and near-infrared photometry of the transit of KIC 1255b would display significantly
smaller depths than those displayed in the optical. Determining that the transit depth of 
KIC 1255b is wavelength dependent, with smaller depths in the near-infrared than the optical, would therefore
strongly favour the explanation of scattering from a dust tail with sub-micron sized particles.

Here we
present an assortment of different observations of KIC 1255 that were obtained in order to either
bolster or rule-out the disintegrating low-mass planet scenario.
In addition to the {\it Kepler} photometry that we analyze here,
these various observational data include:
(i) two Canada-France-Hawaii Telescope/Wide-field
InfraRed Camera (CFHT/WIRCam) Ks-band ($\sim$2.15 micron) photometric
detections of the KIC 1255b 
transit with simultaneous {\it Kepler} photometric detections,
(ii) simultaneous
photometric non-detections of the KIC 1255b transit
with the {\it Hubble Space Telescope} Wide Field Camera 3 ({\it HST}/WFC3) F140W and {\it Kepler} photometry, 
(iii) {\it HST}/WFC3 high-angular resolution imaging of KIC 1255 in the 
F555W ($\lambda$$\sim$0.531 $\mu m$),
and F775W ($\lambda$$\sim$0.765 $\mu m$) bands,
(iv) Keck/NIRC2 ground-based Adaptive Optics (tip/tilt only) high-angular resolution imaging of KIC 1255 in the K'-band ($\lambda\sim$2.124 $\mu m$),
and (v) and Keck/HIRES radial velocity observations of KIC 1255.
The high angular resolution imaging observations
allow us to rule-out nearby background/foreground companions as close as
0.2\arcsec \ to KIC 1255.
Our KECK/HIRES radial velocity observations allow us to rule-out low-mass stellar companions
($\sim$ 0.2 M$_\odot$) for orbital periods $\lesssim 10$ years.
This significantly reduces the parameter space for 
nearby companions to KIC 1255, 
and therefore reduces the odds that the unique 
{\it Kepler} photometry that \citet{Rappaport12} reported
is due to a binary 
or higher-order multiple,
masquerading as a planetary false positive.
Our simultaneous {\it Kepler} and near-infrared detections of the transit of KIC 1255b
appear to report similar depths; as a result, if the 
source of the photometry we observe is a dust tail
trailing a disintegrating planet composed of single-sized particles, then the particles are at least
$\sim$\ParticleSizeCare \ microns in radius. Particles this large can likely only be lofted from a low-mass planet, suggesting
that KIC 1255b might best be described as a sub-Mercury mass planet.

\section{Observations and Analysis}

\subsection{{\it Kepler} photometry}
\label{SecKepler}

\begin{figure}
\centering
\includegraphics[scale=0.47, angle = 270]{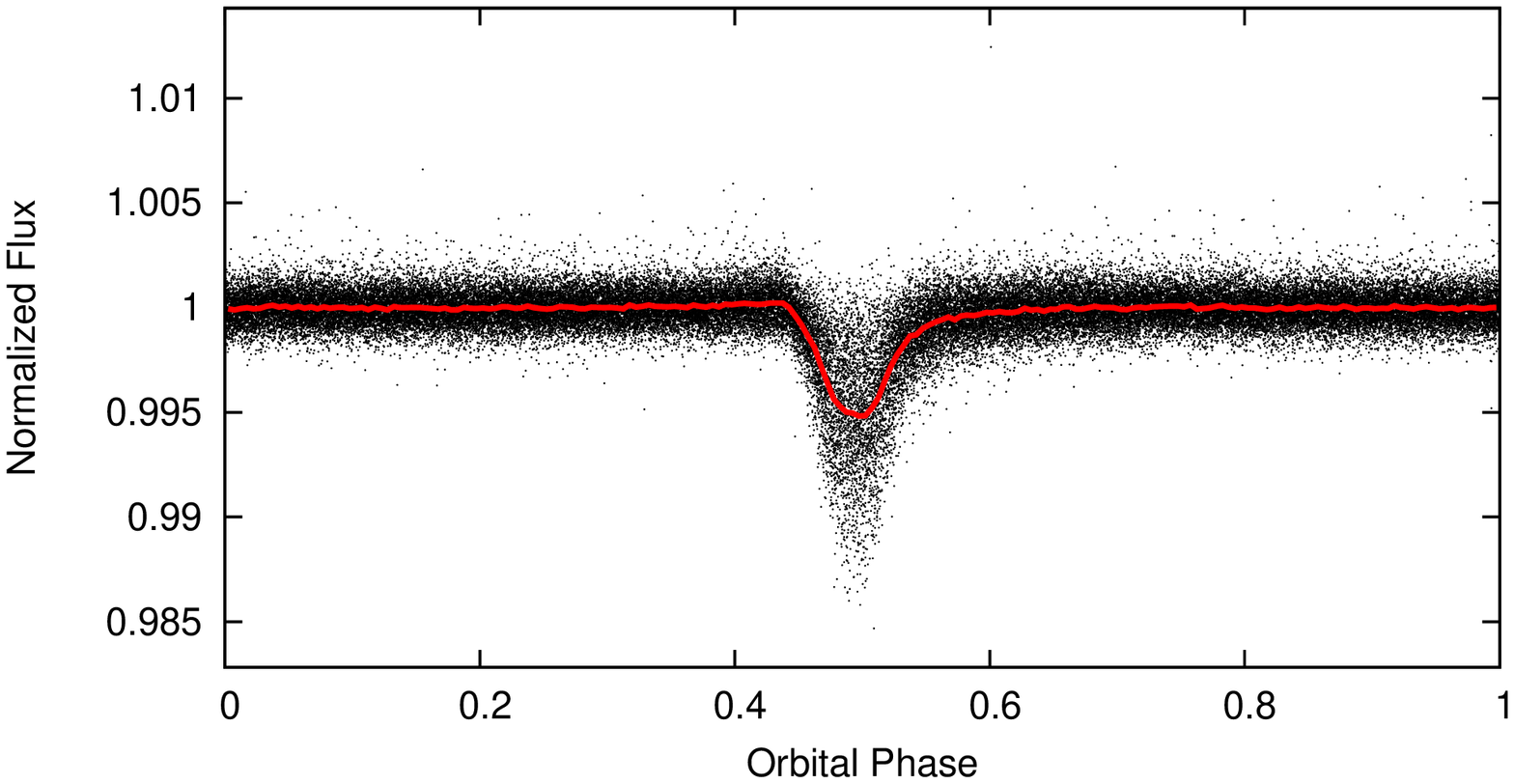}
\includegraphics[scale=0.47, angle = 270]{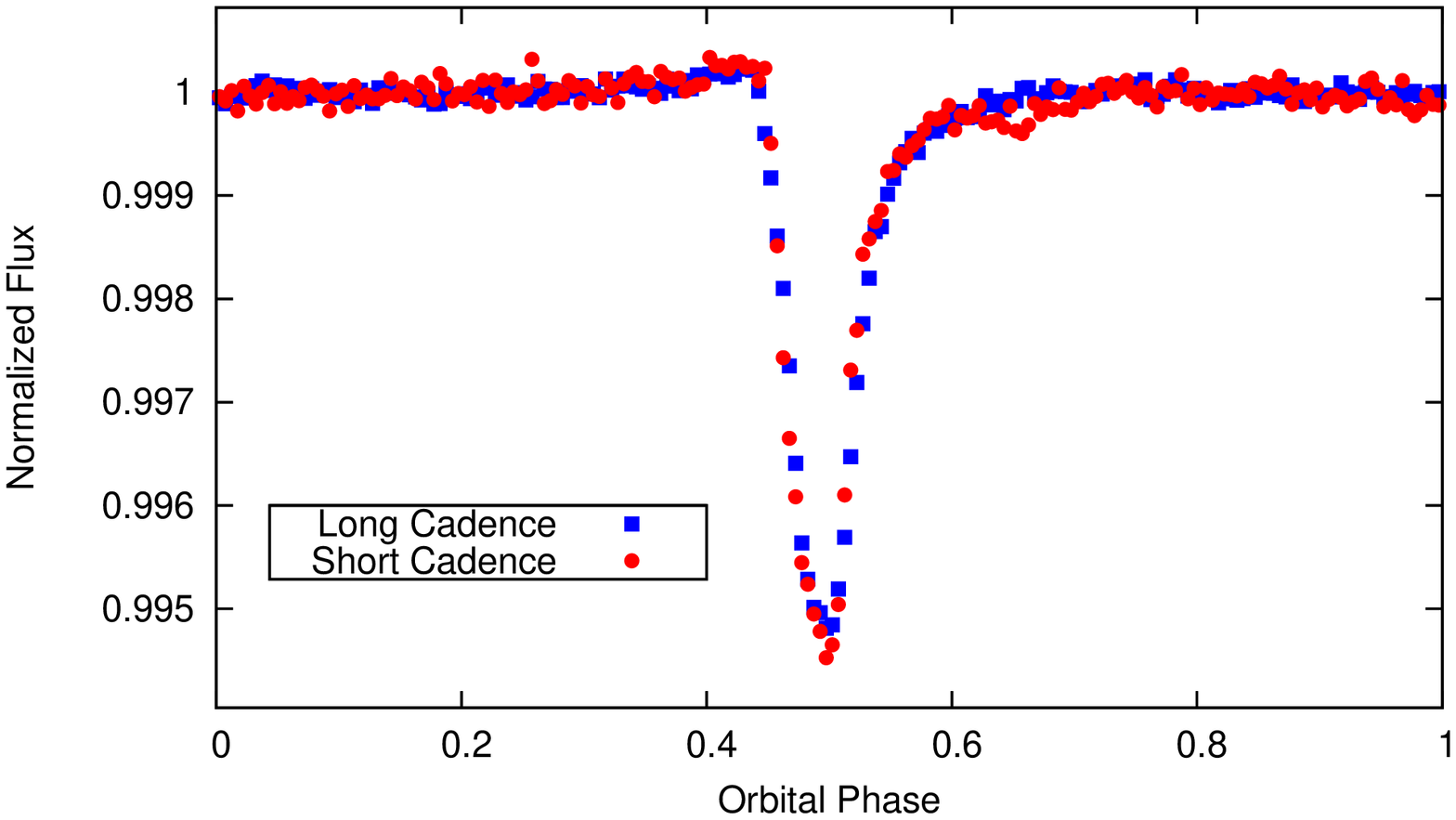}
\caption[KICKepler]
	{
		Top panel: {\it Kepler} long cadence photometry (black points) of KIC 1255 phased to the orbital period
		of the candidate planet ($\sim$15.685 hours). The red line is the binned mean 
		of the orbital phase-folded light curve. 
		Bottom panel: The binned mean of the phase-folded light curve (every
		$\phi$=0.005 in phase) of the short cadence (red circles)
		and long cadence (blue squares) photometry.
	}
\label{FigKeplerPhase}
\end{figure}

%EMULATEAPJCHANGE
\begin{figure*}
%\begin{figure}
\centering
%EMULATEAPJCHANGE, 0.40 for aastex versus 0.44 for emulate apJ
\includegraphics[scale=0.44, angle = 270]{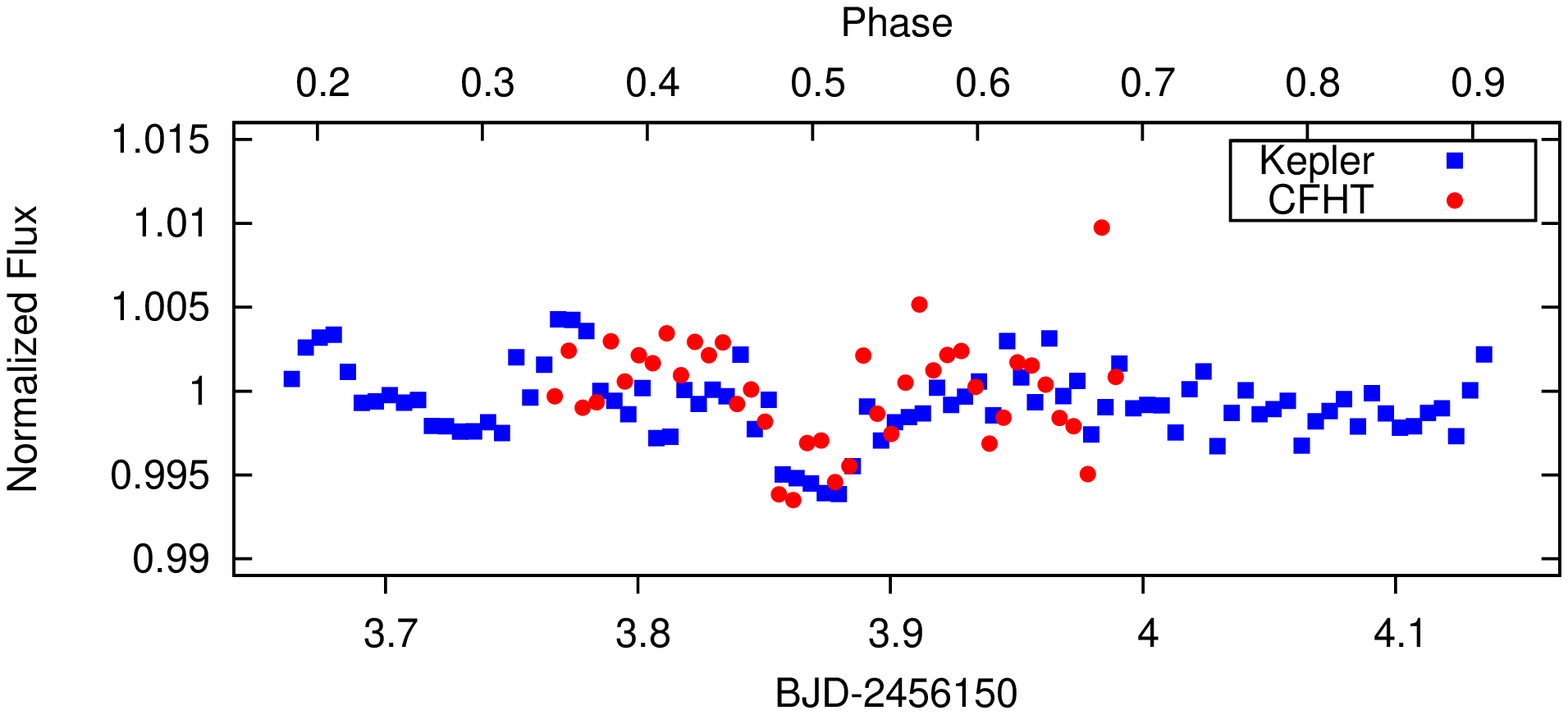}
\includegraphics[scale=0.44, angle = 270]{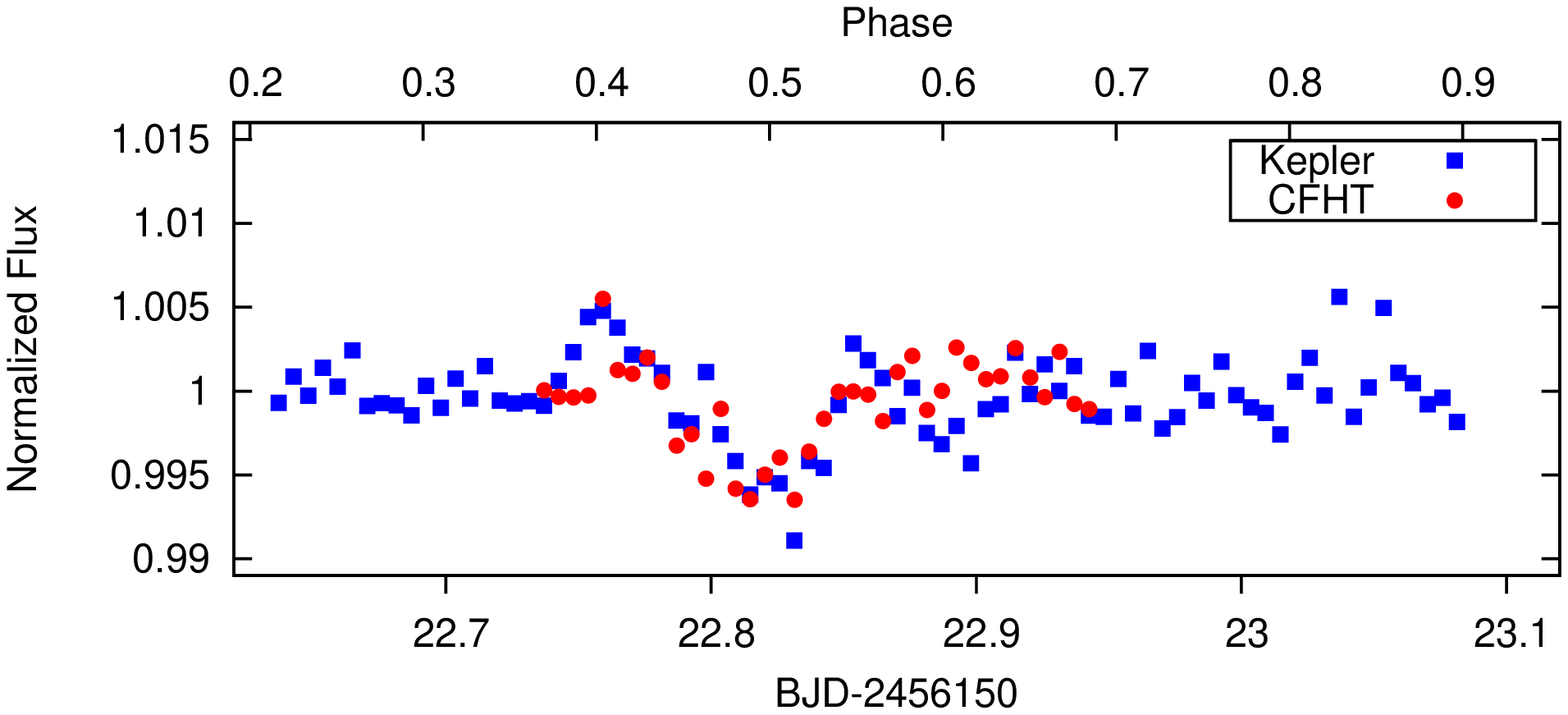}
\includegraphics[scale=0.44, angle = 270]{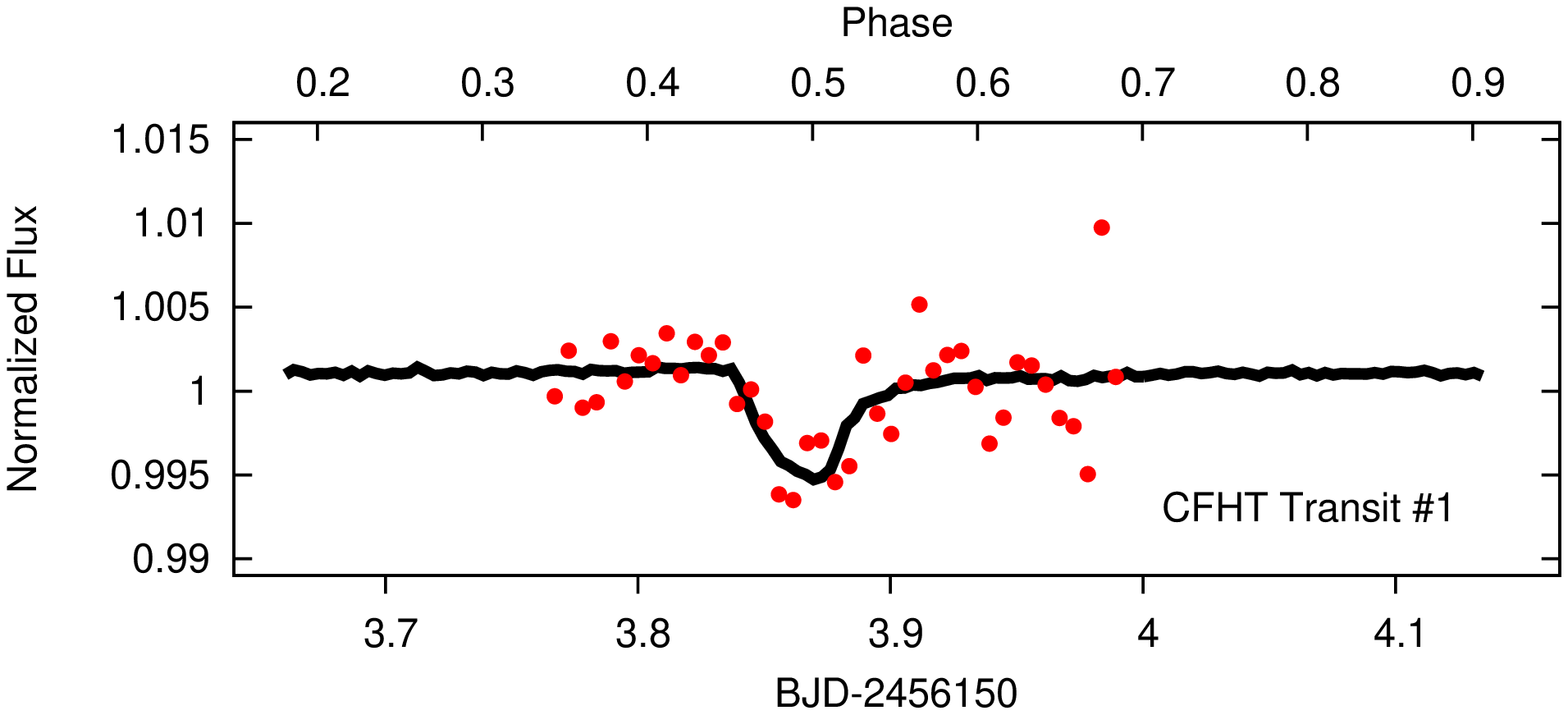}
\includegraphics[scale=0.44, angle = 270]{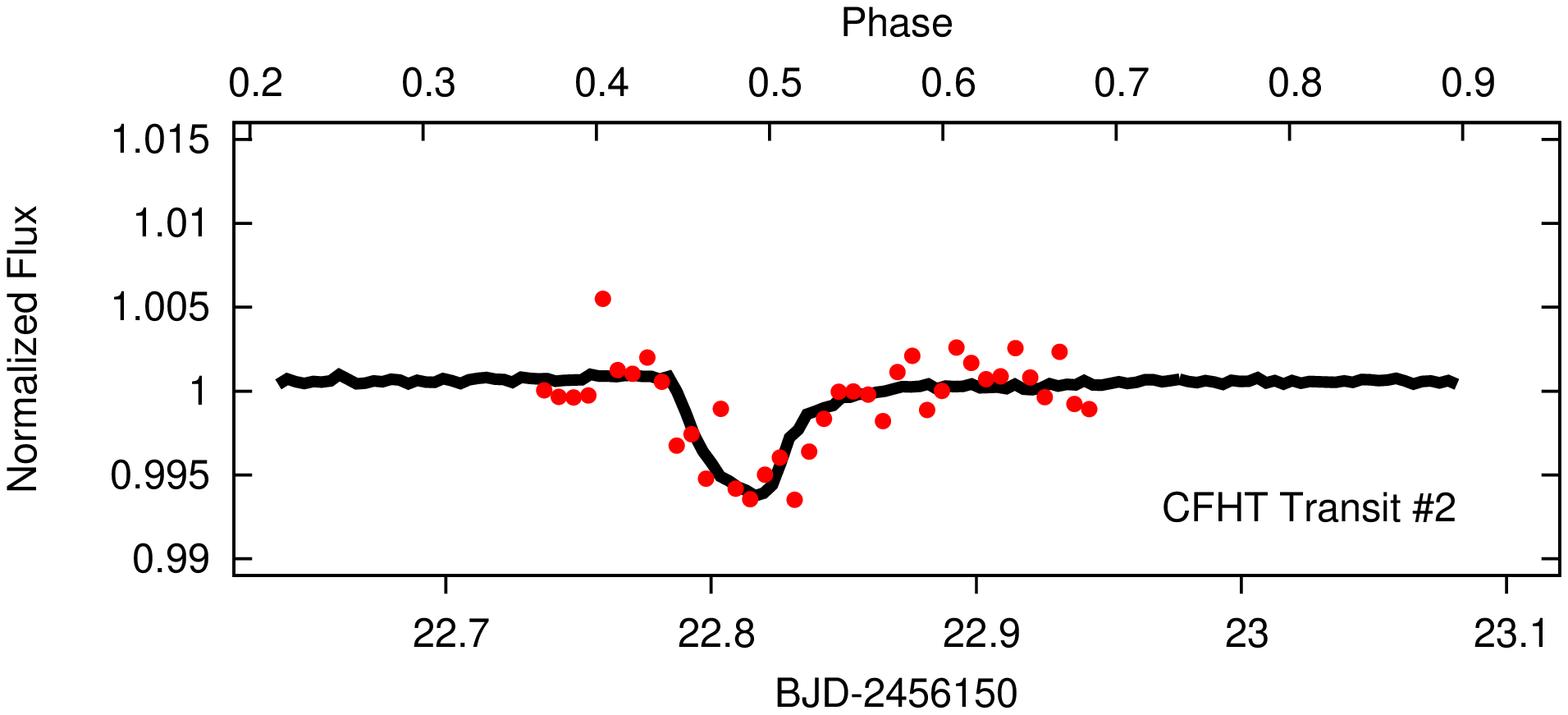}
\includegraphics[scale=0.44, angle = 270]{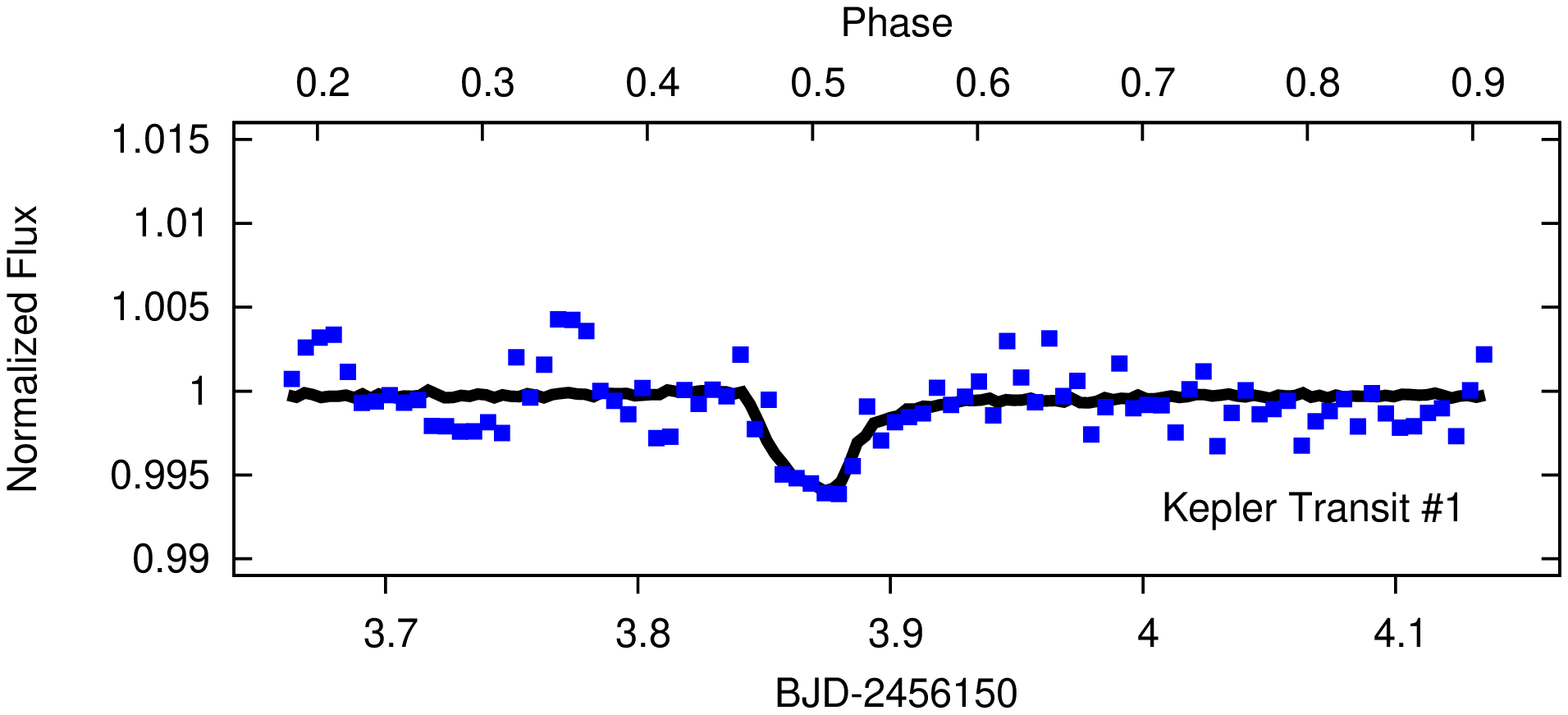}
\includegraphics[scale=0.44, angle = 270]{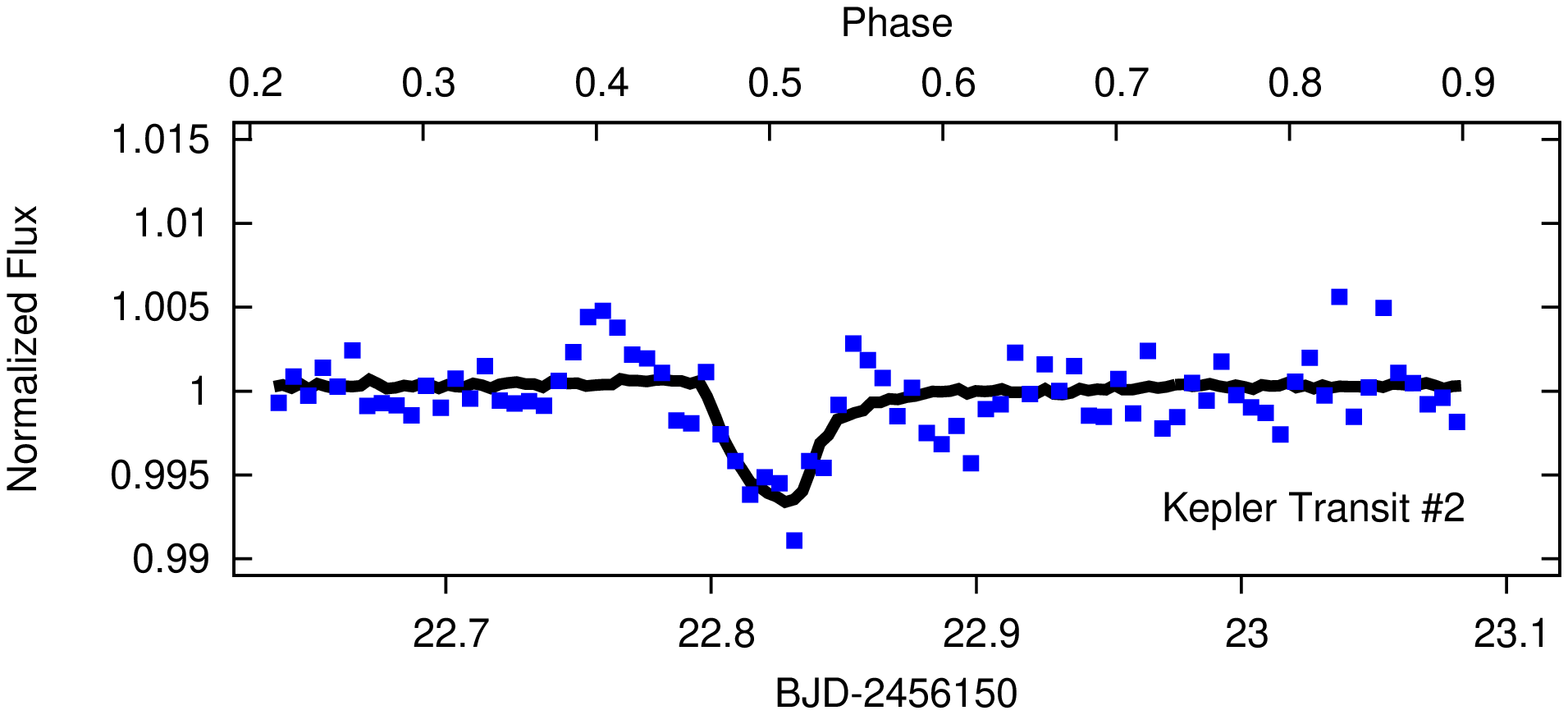}
\caption[KICCFHT]
	{
		CFHT/WIRCam Ks-band photometry (red circles) and {\it Kepler} photometry (blue squares)
		of the transit of KIC 1255b obtained on the nights
		of 2012 August 13 (left panels) and 2012 September 1 (right panels; Hawaiian Standard Time).
		Both sets of data are binned every 8 minutes.
		The black lines in the bottom two sets of panels indicate the
		best-fit transit model for each set of data; the transit model is a scaled version
		of the mean of the short-cadence, phase-folded {\it Kepler} light curve ($g(t)$; see $\S$\ref{SecKepler}).
	}\label{FigCFHT}
%EMULATEAPJCHANGE
\end{figure*}
%\end{figure}

\begin{figure}
\centering
\includegraphics[scale=0.42, angle = 270]{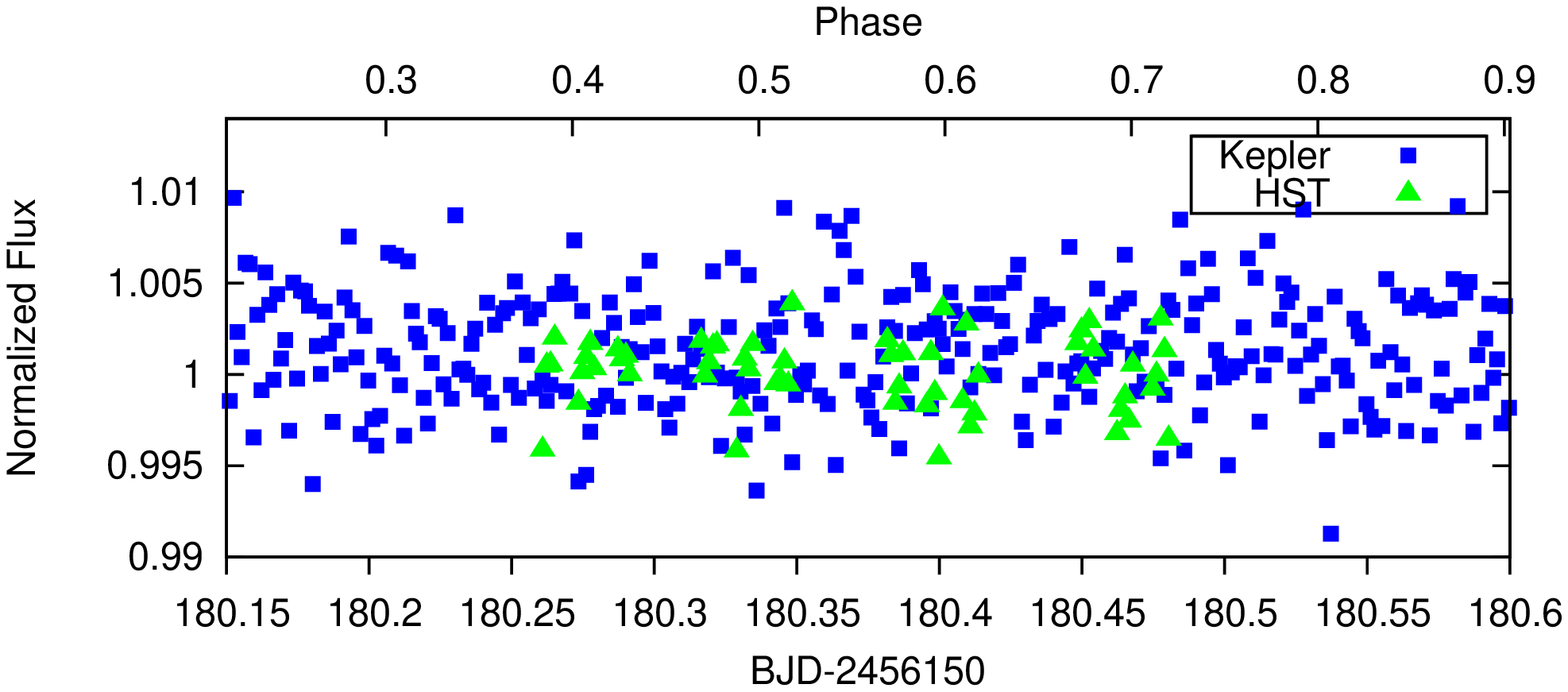}
\includegraphics[scale=0.42, angle = 270]{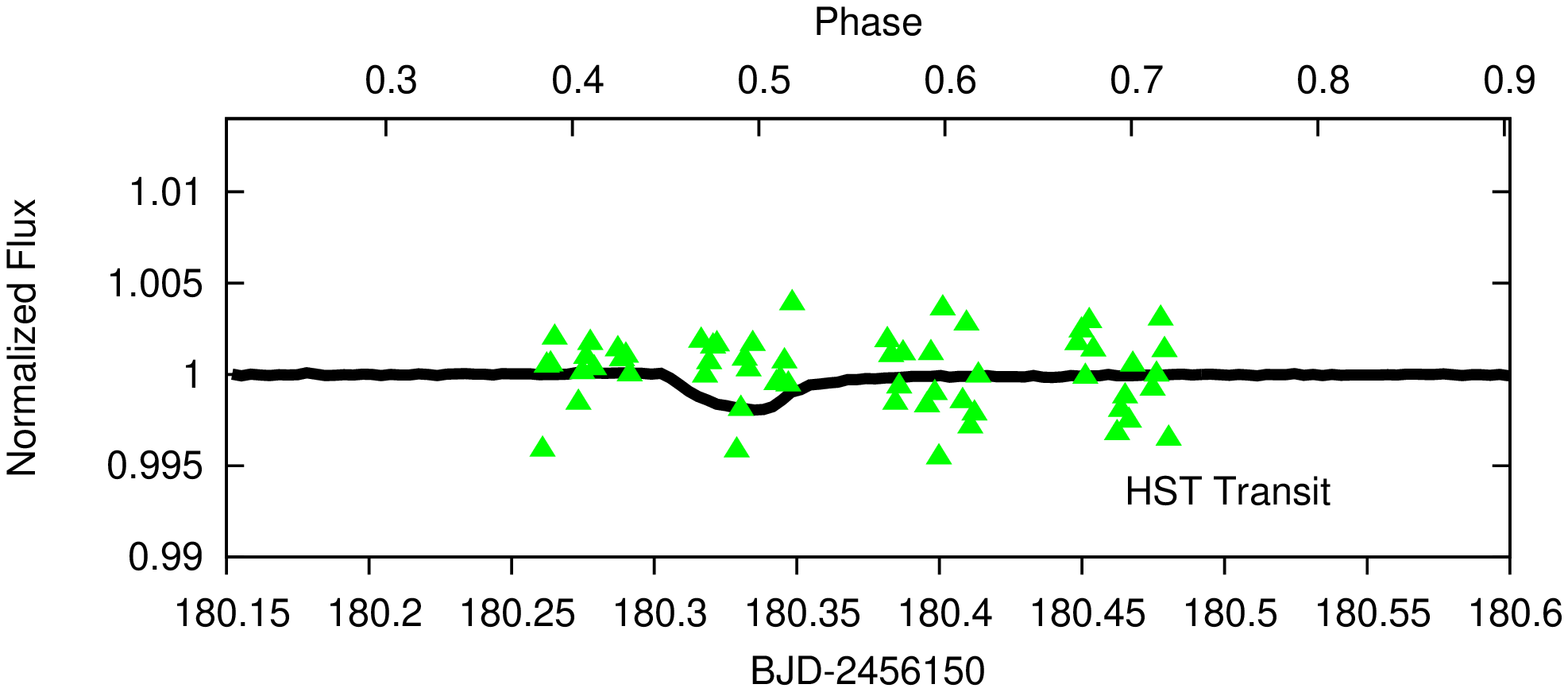}
\includegraphics[scale=0.42, angle = 270]{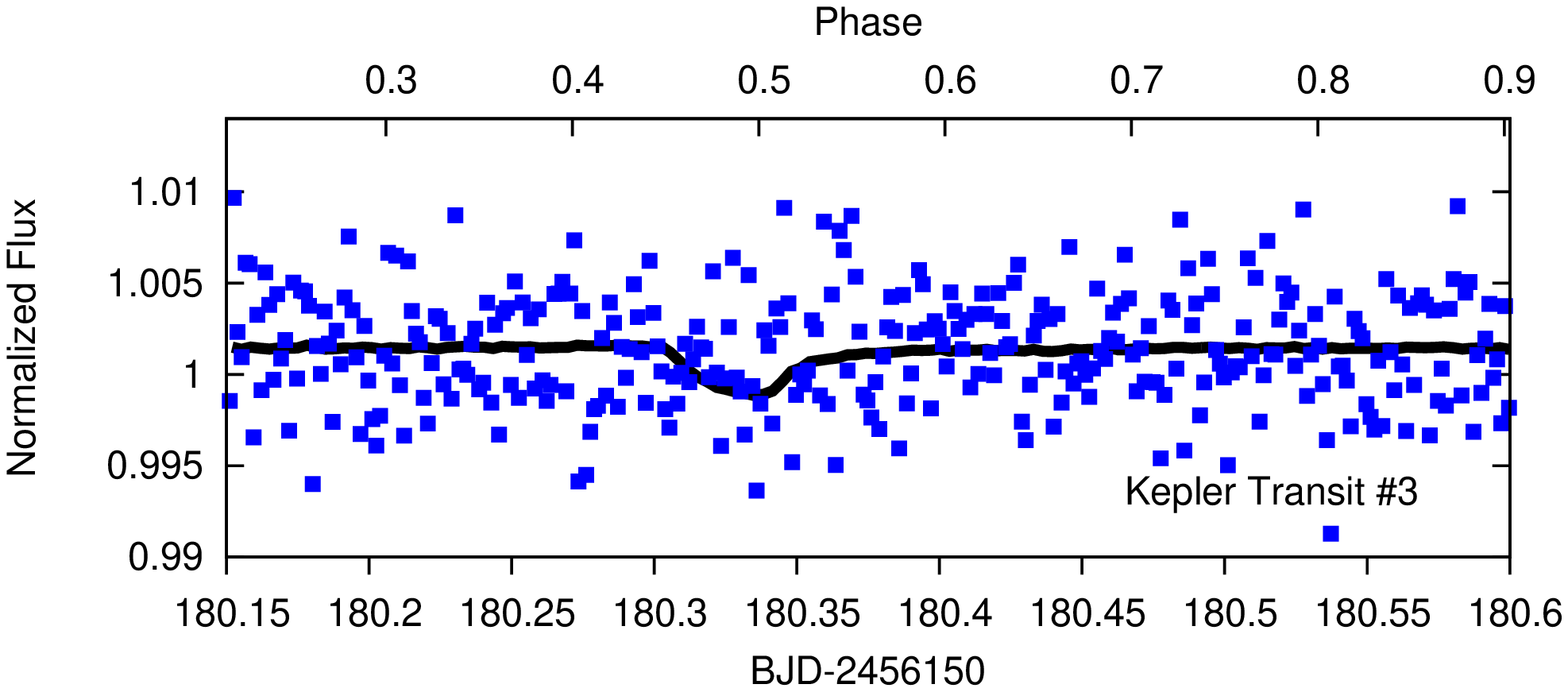}
\caption[KICHSTPhot]
	{
	{\it HST} F140W filter photometry (1.4$\mu m$; green triangles) and {\it Kepler} photometry (blue squares)
	of the transit of KIC 1255b obtained on 2013 February 6 (UTC).
	%The {\it Kepler} data is binned every 2 minutes, while the {\it HST} is binned every 2 minutes.
	The {\it Kepler} and the {\it HST} data are binned every 2 minutes.
	The black lines in the bottom two panels indicate the 3$\sigma$ upper limit on the transit depth
	obtained by scaling the mean of the short-cadence, phase-folded {\it Kepler} light curve ($g(t)$; see $\S$\ref{SecKepler}).
	}
\label{FigHSTPhotometry}
\end{figure}

We start by presenting an analysis of the {\it Kepler} data of KIC 1255. We analyze all the
long cadence (quarters 1-16 at the time of writing; 29.4 minute sampling)
and short cadence (quarters 13-16; 58.8 second sampling) Simple Aperture Photomery (SAP) of this star.
KIC 1255 displays obvious rotational modulation with a period of $\sim$22.9 $d$ that varies 
between 1-4\% of the observed stellar flux; we remove this modulation 
by employing a cubic spline. To ensure that KIC 1255b's variable and asymmetric transit depth in no way 
impacts our starspot removal technique, we cut out all data in the transit before calculating our cubic spline;
that is, we phase the data to the orbital period
of the putative planet (where phase, $\phi$=0.5 denotes the midpoint of the transit), and cut out all data between
phases $\phi$=0.4-0.7\footnote{The asymmetric cut around the mid-point of the transit is obviously due to the asymmetric
shape of the transit.}, and then bin the data every $\sim$10 hours.
% In this way we ensure the broad, asymmetric transit depth does not influence our spot removal technique.
After calculating the cubic spline on the binned {\it Kepler} data with the transits removed,
we apply the cubic spline to all the unbinned {\it Kepler} data and apply a 10$\sigma$ cut to remove outliers; we thus produce 
a light curve of KIC 1255 with the obvious rotational modulation removed. The spot-corrected, phase-folded, long cadence
light curve of KIC 1255 is presented in the top panel of Figure \ref{FigKeplerPhase}; the short-cadence data are similar,
but have much higher scatter per point.
We present the phase-folded, binned mean of the long 
and short cadence data in the bottom panel of Figure \ref{FigKeplerPhase}.
We note that the short cadence data display a marginally 
narrower transit,
and appears to have  an extra,
brief, enhanced decrement in flux following the transit (i.e., near phase
$\phi$=0.65) that is not visible in the long cadence photometry

% and have an obvious decrement in flux following the transit that is not visible in the long cadence photometry.

 To compare the {\it Kepler} photometry
to our CFHT groundbased ($\S$\ref{SecCFHT})
and {\it HST} spacebased ($\S$\ref{SecHSTPhotometry})
photometry, we also present the {\it Kepler} SAP photometry,
without the spline-correction. 
Given the asymmetric transit profile and varying transit depth displayed in the {\it Kepler} photometry,
we choose to fit our individual {\it Kepler} transits
(and the simultaneous CFHT and {\it HST} photometry)
by scaling the mean transit profile of the short cadence {\it Kepler} photometry 
by a multiplicative factor, $A$.  
Therefore the function we use to fit our data, $g(t)$ is compared to the 
mean of the phase-folded  short cadence photometry, $f(t)$ (shown in the bottom panel of Figure \ref{FigKeplerPhase}), by:
\begin{equation}
g(t) =  1 + A \,[ f(t-t_{transit}) - 1 ] + y_o
\label{EquationFtGt}
\end{equation}
where $y_o$ is simply a vertical offset,
and $t_{transit}$ is the mid-transit time (defined as the minimum of the phase-folded mean of the short cadence photometry,
at phase $\phi$=0.5).
We note that a value of $A$ = 1 corresponds to a KIC 1255b transit depth of 0.55\% of the stellar
flux, as shown in the bottom panel of Figure \ref{FigKeplerPhase}.
We note that by multiplying our
phase-folded mean by $A$ we are scaling up or down the size of the apparent forward-scattering peak, as well as the depth
of the transit\footnote{This assumption is likely reasonable, as 
the analysis of \citet{Brogi12} indicates that the deeper transits appear to display a larger 
forward scattering peak just prior to transit, as one might naively expect if the deeper transit is being caused
by a larger amount of material occulting the star.}.
To fit our {\it Kepler} transits, as well as the CFHT and {\it HST} transits that
follow\footnote{We note that we do not account for the effect of the various exposure times of our
CFHT, {\it HST} and {\it Kepler} data on estimating the parameters of interest in Equation \ref{EquationFtGt}, as such differences
are negligible given our short exposure times ($\sim$5 seconds to $\sim$1 minute; please see \citealt{Kipping10}).}, 
we employ Markov Chain Monte Carlo (MCMC) techniques (as described for our purposes in \citealt{CrollMCMC}).
In Figures \ref{FigCFHT} and \ref{FigHSTPhotometry} we present the SAP
short cadence {\it Kepler} photometry of KIC 1255b that was obtained simultaneously 
with the CFHT and {\it HST} photometry, as well as the best-fit scaled profile, $g(t)$,
of the mean short-cadence {\it Kepler} profile, $f(t)$.
We assume an error on the {\it Kepler} data for our MCMC fitting based on the RMS of the residuals to the best-fit model.
The associated best-fit parameters are presented in Table \ref{TableParams}.

\subsection{CFHT/WIRCam photometry}
\label{SecCFHT}

%EMULATEAPJCHANGE
\begin{figure*}
%\begin{figure}
\centering
%EMULATEAPJCHANGE, 0.40 for aastex, 0.44 for emulateapj
\includegraphics[scale=0.44, angle = 270]{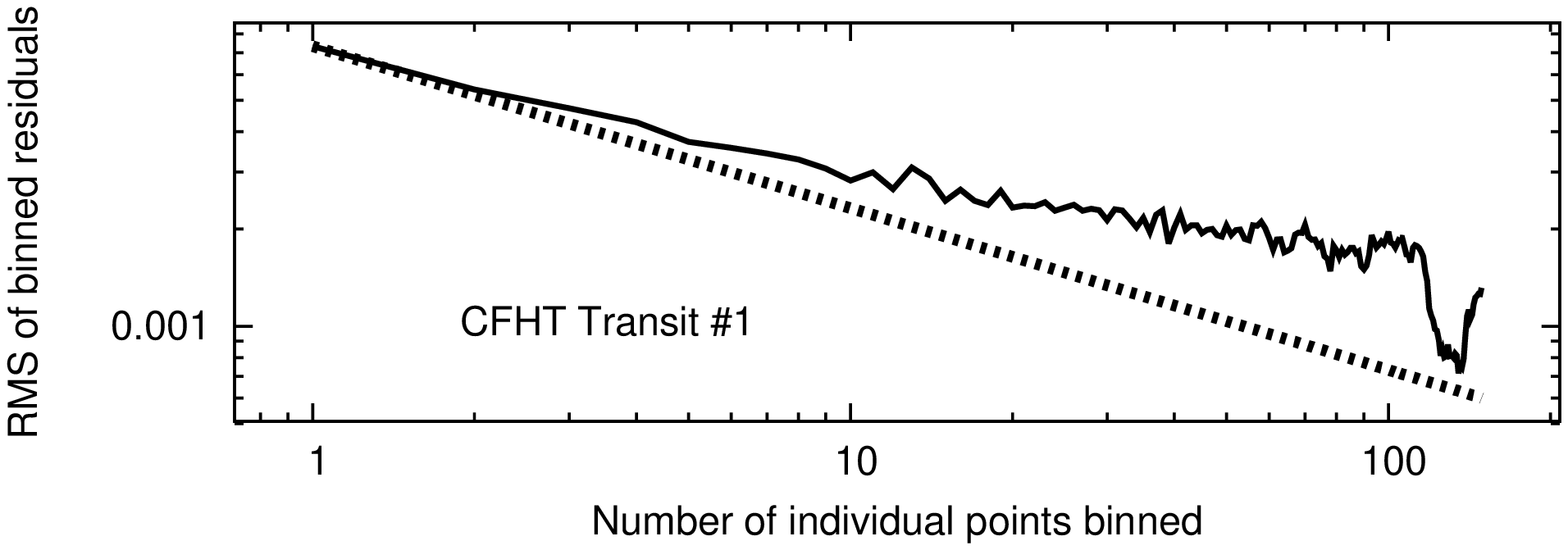}
\includegraphics[scale=0.44, angle = 270]{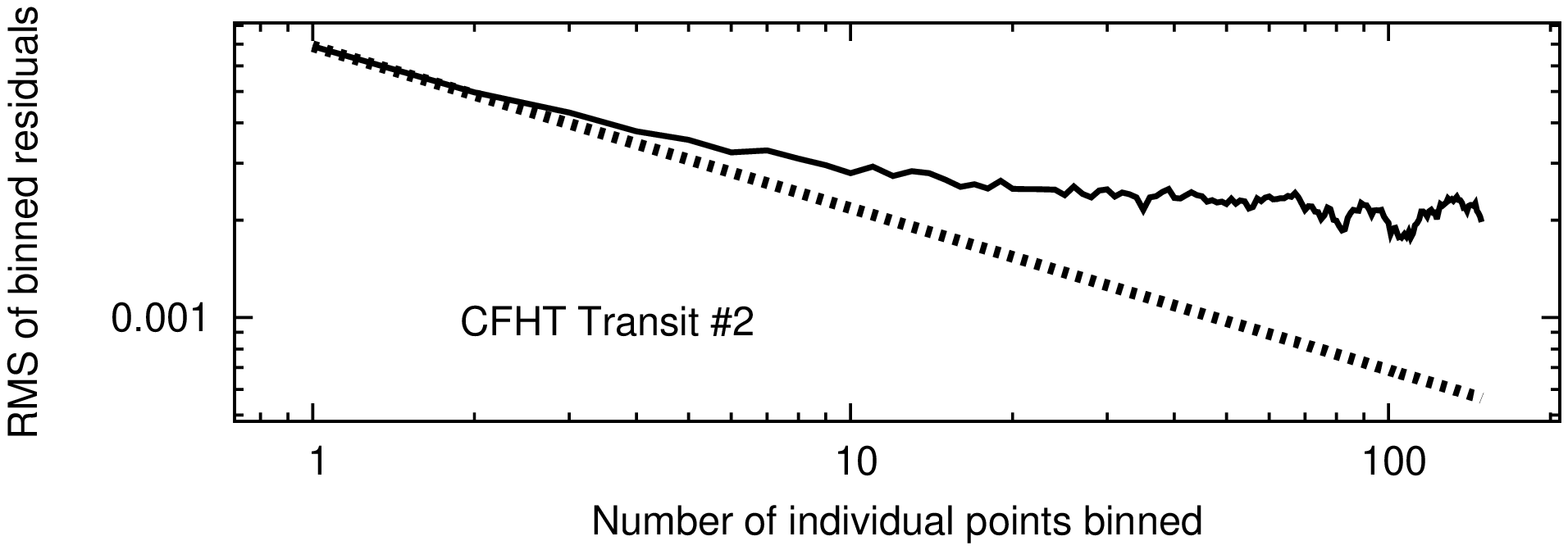}
\caption{	The root-mean-square of our photometry after the subtraction of our best-fit (solid line)
		for our CFHT/WIRCam Ks-band photometry of our first-transit
		(left), and our second-transit (right).
		The dashed line in each panel displays the
		one over the square-root of the bin-size expectation for Gaussian noise.
		}
\label{FigPoisson}
%EMULATEAPJCHANGE
\end{figure*}
%\end{figure}

 We obtained two Ks-band ($\sim$2.15 micron) WIRCam \citep{Puget04} photometric data-sets of the transit 
of KIC 1255 ($K$$\sim$13.3). 
Data-sets were obtained on 
the evenings of 2012 August 13 and 2012 September 1 (Hawaiian Standard Time).
The observations lasted for $\sim$6.5 hours and $\sim$5 hours, respectively. 
High wind impacted the image quality for the first set of observations (2012 August 13); the second 
set of observations were of photometric quality throughout the night (2012 September 1).
%Both occasions were photometric 
%, other than a passing cloud which may have briefly affected the photometry around BJD$\sim$=
Reduction of the data and aperture photometry was performed as detailed in
\citet{CrollTrESTwo,CrollTrESThree}.
Although WIRCam offers a 21\arcmin$\times$21\arcmin \ field-of-view, we only utilize reference stars
from the same detector as our target, therefore resulting in a 10\arcmin$\times$10\arcmin \ field-of-view.
We employ a range of aperture radii for our CFHT photometry (as discussed below in $\S$\ref{SecCorrelation}),
and subtracted the sky in all cases using
an annulus with an inner and outer radius of 14 and 20 pixels, respectively.
To determine the fractional contribution of the square 
pixels at the edge of the circular aperture we multiply the 
flux of these pixels by the fraction of the pixels that falls within our circular aperture;
we determine this fractional contribution using the 
GSFC Astronomy Library IDL procedure {\it pixwt.pro}.
The exposure times were 25 seconds, with an overhead for read-out and 
saving exposures of 7.38 seconds,
resulting in an overall duty-cycle of $\sim$76\%.
The telescope was defocused to 0.9 mm for both observations.
In both cases at the conclusion of the observations the airmass increased
to beyond 2.0; we noticed reduced precision in the resulting light curves
once the airmass rose above a value of 1.6. As a result, we
exclude all data with an airmass greater than this latter value in the following analysis.
Our CFHT photometry is presented in the middle panels of Figure \ref{FigCFHT},
using aperture radii of 7 and 9 pixels for our first and second CFHT transits, respectively.
After the subtraction of our best-fit models,
we achieve an RMS precision of 7.1$\times$$10^{-3}$ and 6.5$\times$$10^{-3}$ per exposure
for our first and second transit, respectively. % and readout (32.38 seconds).
This compares to the expected photon noise limit of 1.15$\times$$10^{-3}$ per exposure, or
3.52$\times$$10^{-3}$ once other noise sources (read-noise, dark and sky-noise, where the sky is the dominant component)
are taken into account.
We compare our photometric precision to the Gaussian noise expectation of
one over the square-root of the bin-size
in Figure \ref{FigPoisson}. Both data-sets scale down above this limit, indicative of correlated noise;
some of this correlated noise is likely
astrophysical as the {\it Kepler} data display obvious modulation, likely due to
granulation or evolution of starspots, plages, etc.,
that appears to be partially replicated in the CFHT near-infrared photometry.

We fit our CFHT Transits with the scaled-version of the 
short-cadence, phase-folded {\it Kepler} light curve ($g(t)$; $\S$\ref{SecKepler}).
The best-fit transits are displayed in Figure \ref{FigCFHT}, and the 
associated transit depths, $A$, compared to the mean {\it Kepler} transit depth, are
given in Table \ref{TableParams}.

We also note that as our CFHT photometry has slightly superior angular resolution
than the coarse Kepler pixels, our CFHT photometry allows us to place a limit on the angular
separation of the transiting object from KIC 1255. 
Assuming the transits we observe in our CFHT photometry are due to the same object
causing the transits we observe in the Kepler data, and given the 
7-9 pixel aperture we use here and WIRCam's 0.3\arcsec pixel-scale,
the object causing the transits we associate with the candidate planet KIC 1255b cannot be due to
a companion more than $\sim$2\arcsec away from KIC 1255.

\subsubsection{Correlations with Aperture size and Number of Reference Stars}
\label{SecCorrelation}

%EMULATEAPJCHANGE
\begin{figure*}
%\begin{figure}
\centering
%EMULATEAPJCHANGE, 0.40 for aastex, 0.43 for emulateapj
\includegraphics[scale=0.43, angle = 270]{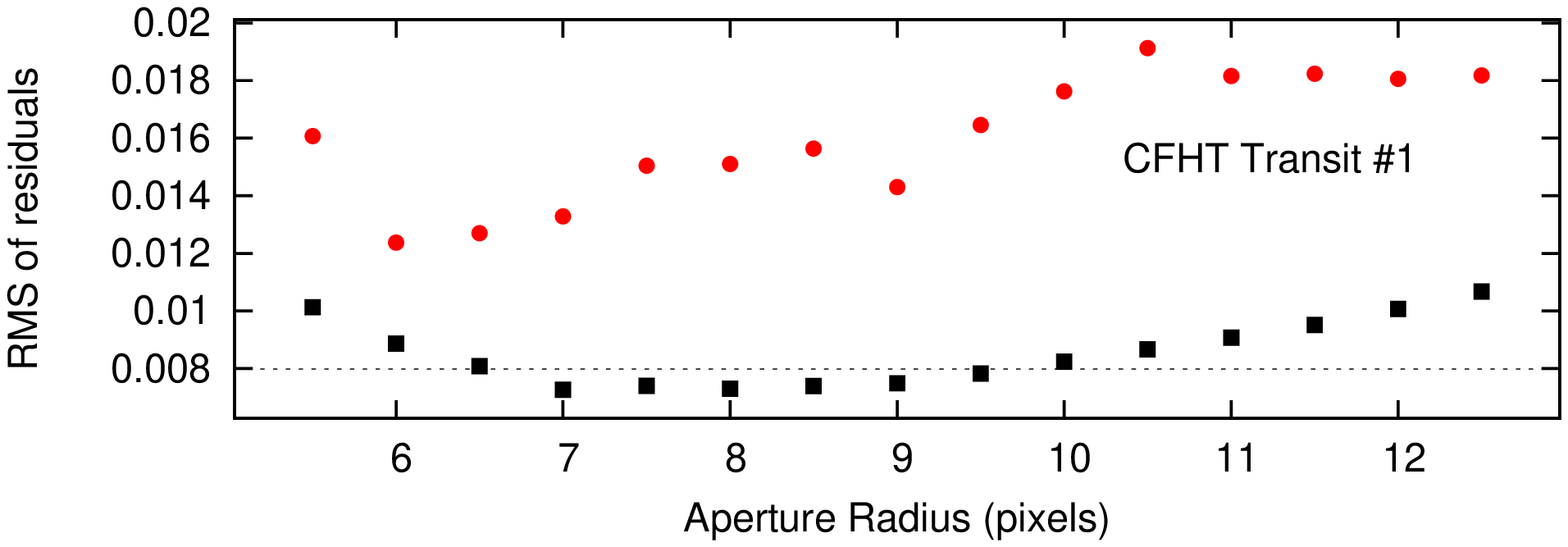}
\includegraphics[scale=0.43, angle = 270]{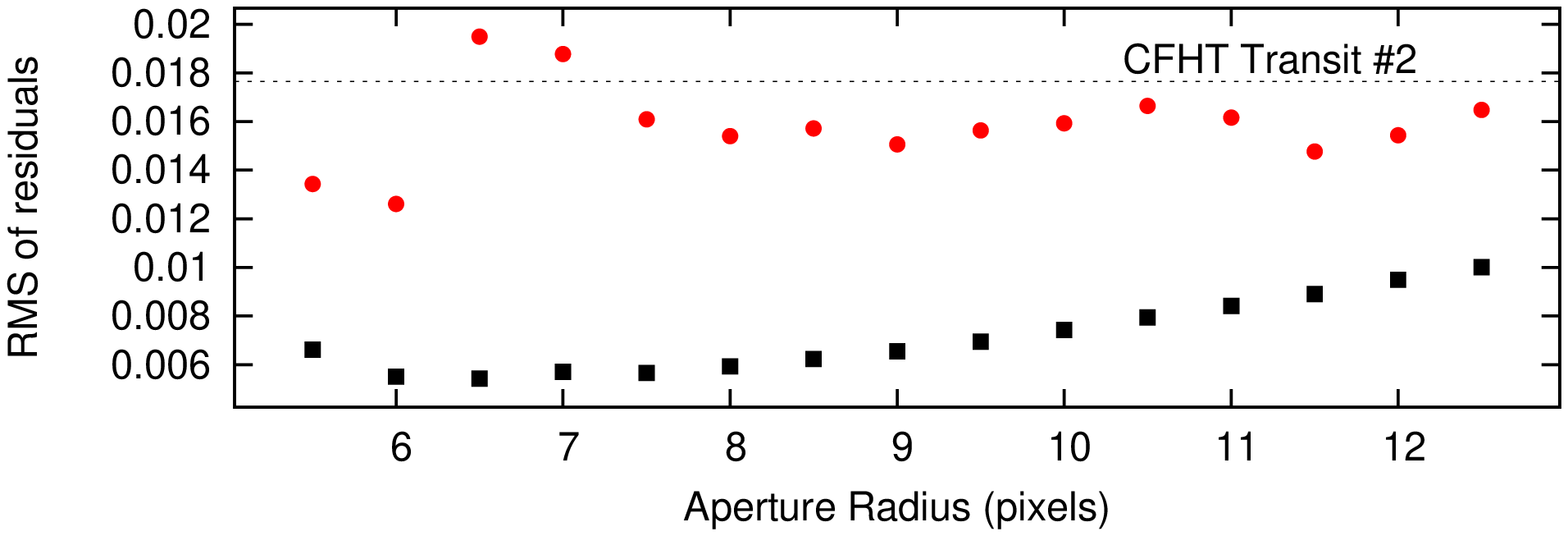}

\includegraphics[scale=0.43, angle = 270]{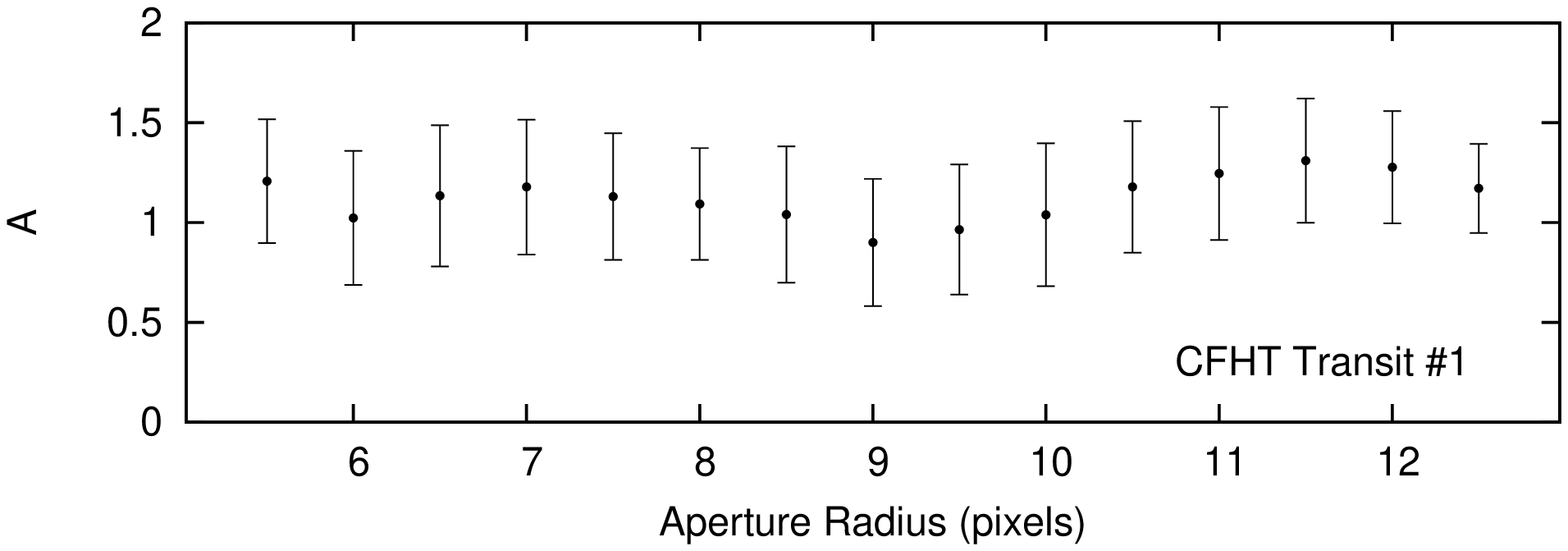}
\includegraphics[scale=0.43, angle = 270]{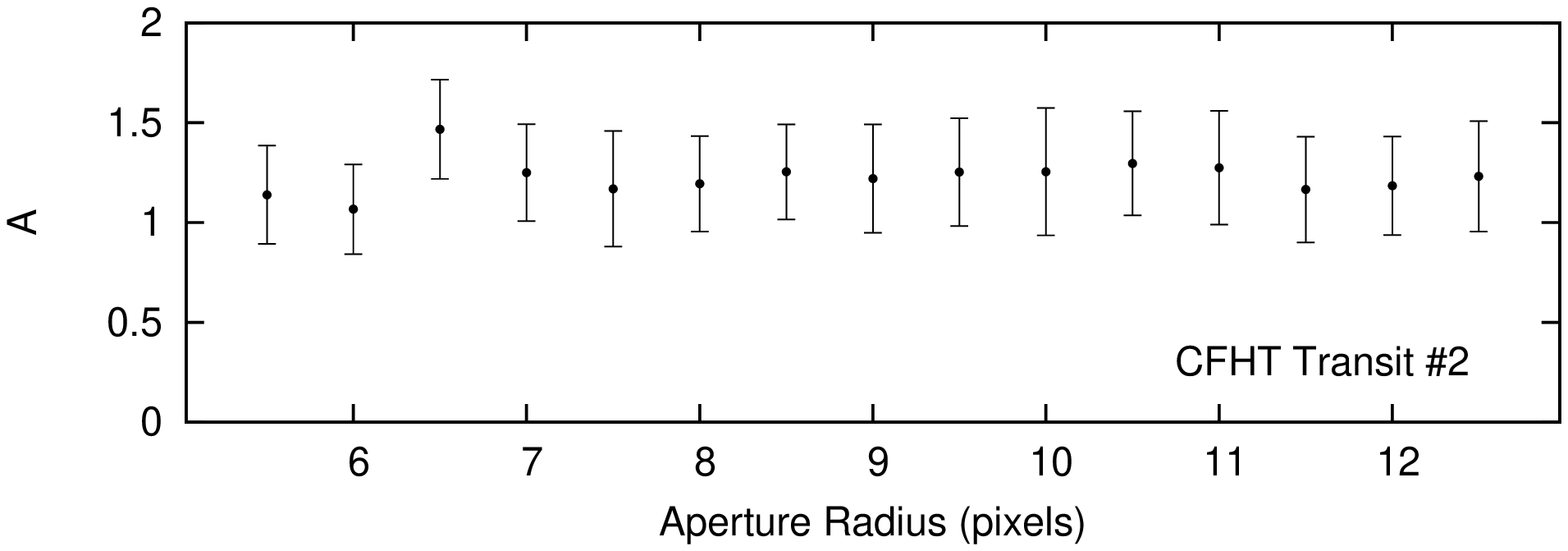}

\includegraphics[scale=0.43, angle = 270]{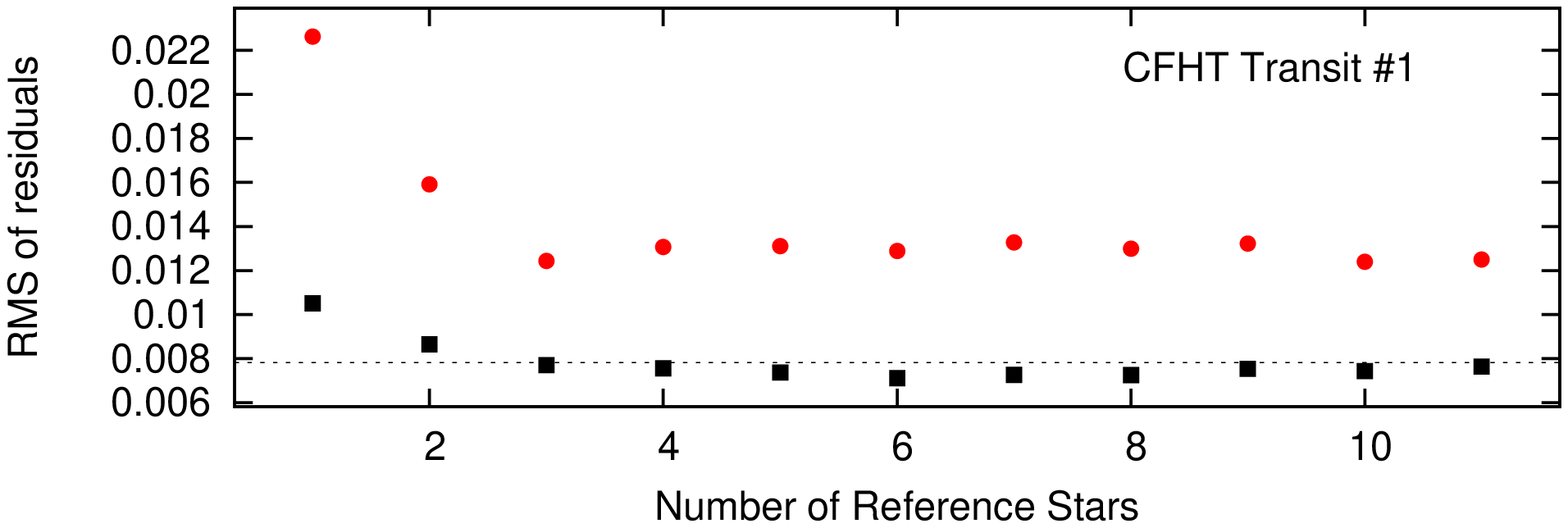}
\includegraphics[scale=0.43, angle = 270]{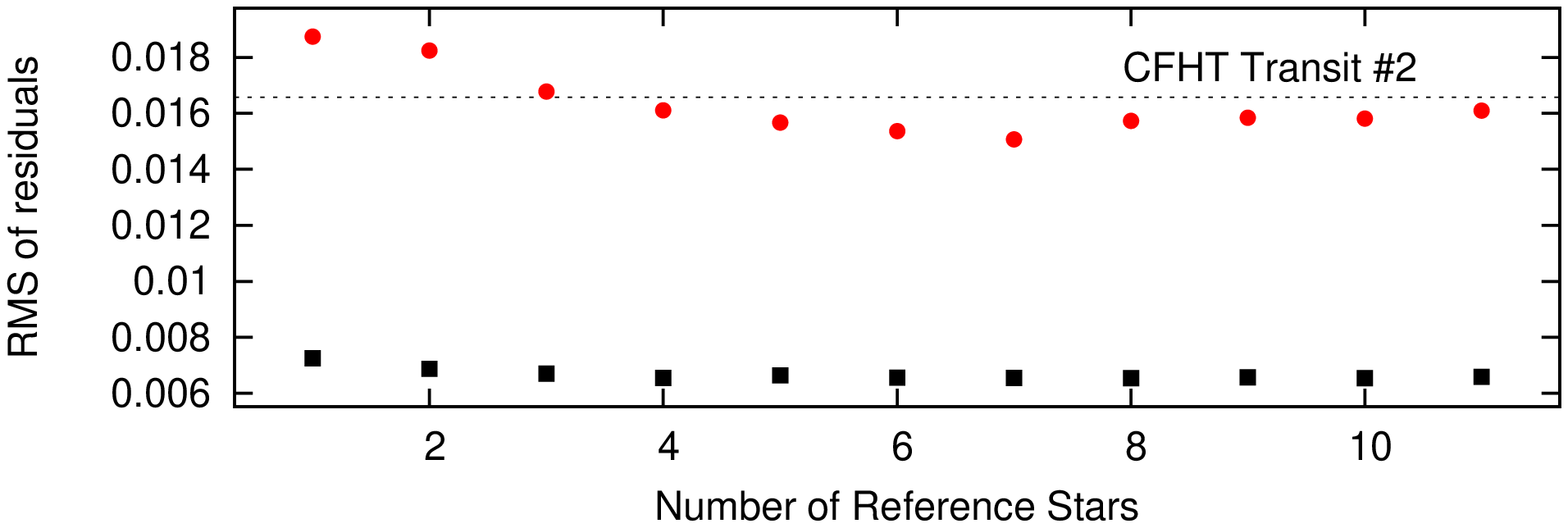}

\includegraphics[scale=0.43, angle = 270]{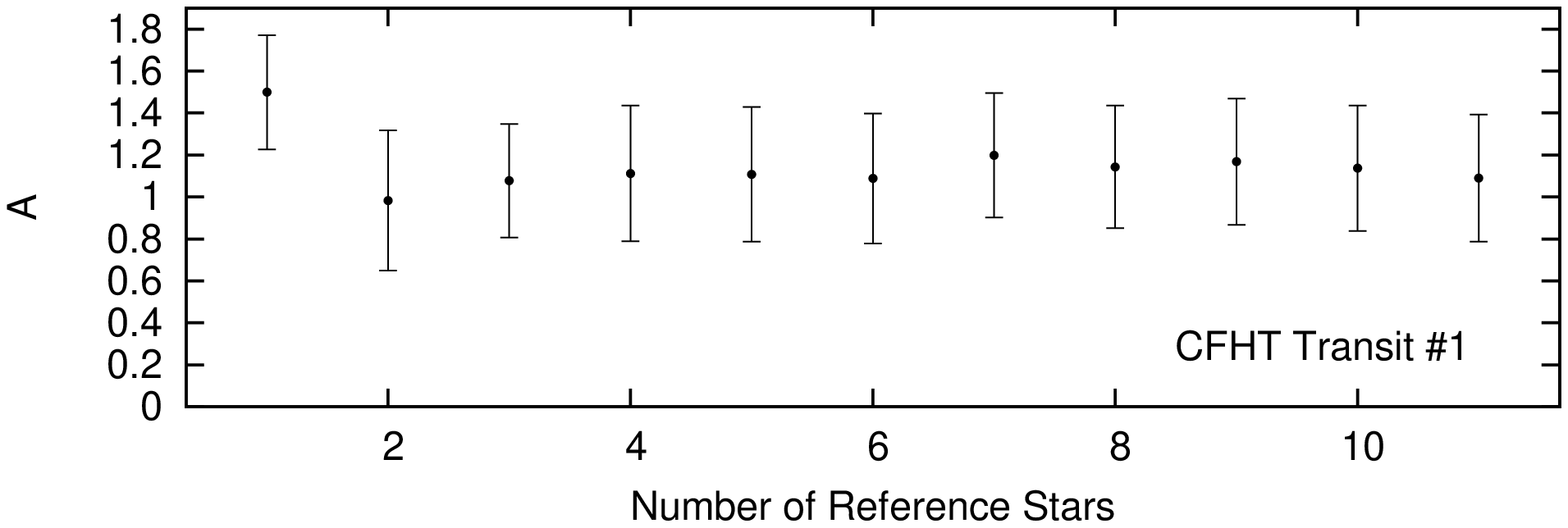}
\includegraphics[scale=0.43, angle = 270]{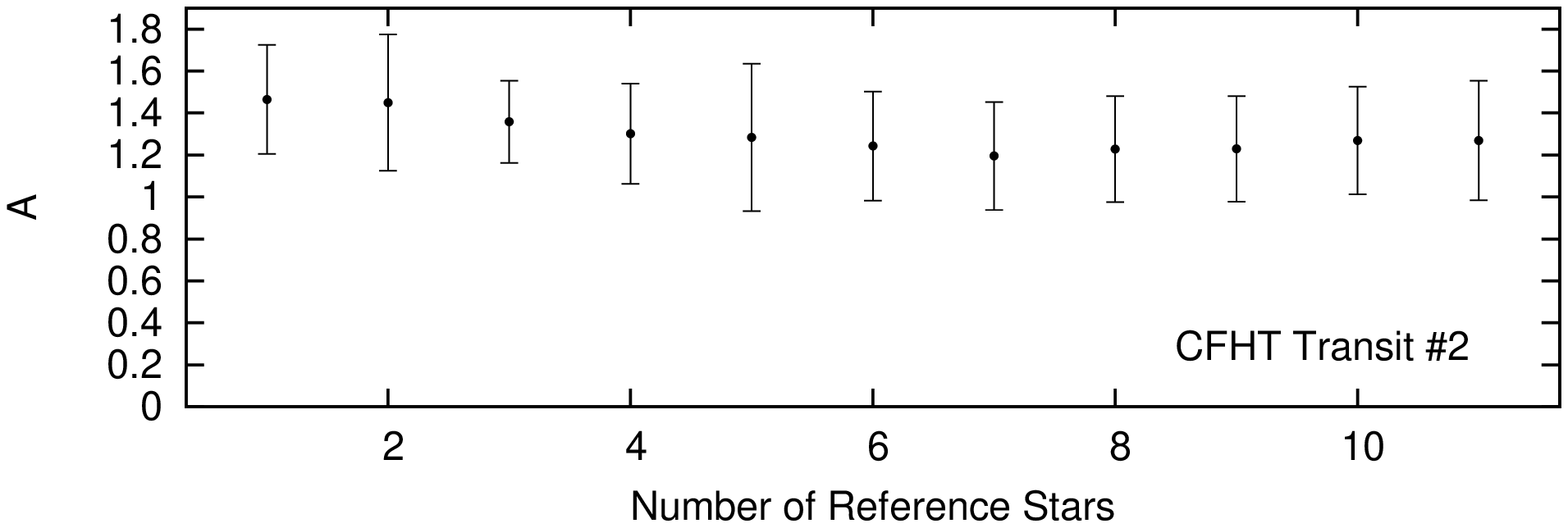}

\caption[KIC1255 Fidelity]
	{
		Top panels: The RMS of the residuals from the best-fit MCMC fit (black squares),
		and the residuals multiplied by the relevant $\beta$ factor (red circles)
		for each of the various aperture sizes (using the best 7 reference stars),
		for our first (left) and second (right) transit CFHT/WIRCam Ks-band transit.
		Second from top set of panels: The associated measured 
		MCMC transit depths, $A$.
		The bottom four panels are the same as the top four, except they
		display the RMS and $A$ values for the different number of 
		reference stars 
		for our first transit (left; 7.0-pixel aperture radius)
		and our second transit (right; 9.0-pixel aperture radius).
		We average over all $A$ values for aperture radii and number of reference stars
		with RMS values (black points) below the black dotted line for the first CFHT transit (left panel);
		for the second transit (right panel) we average all $A$ values with
		RMS$\times$$\beta$ values (red points) below the black dotted line.
		}
\label{FigKICFidelity}
%EMULATEAPJCHANGE
\end{figure*}
%\end{figure}

We also search for correlations in our CFHT photometry between the measured transit 
depth and our choice of aperture size, and the choice of the number of reference stars.
With our other CFHT photometry of hot Jupiters (Croll et al. in prep.),
in some cases we noticed moderate correlations between the secondary eclipse or primary
transit depths and the aperture radius chosen for 
aperture photometry, or the choice of the ensemble of reference stars we use to correct our photometry.
Despite these changes
in the transit/eclipse depth, the differences 
in the RMS of the residuals to the best-fit model were often negligible. Therefore we were confronted with a range
of seemingly equally good fits to the data, where, troublingly,
the parameter of interest, the transit/eclipse depth, varied significantly.
As a result, rather than quoting just the best-fit
of a single aperture and reference star combination, we quote the weighted mean of a number of aperture photometry and reference
star combinations, and scale up the associated error to take into account these correlations.

To decide the best reference star and aperture radius combination, we generally use the RMS of the residuals
multiplied by $\beta$, which parameterizes the amount of time-correlated red-noise in the 
photometry. $\beta$ is defined, employing the methods of \citet{Winn08}, as the factor by which the residuals scale above
the Gaussian noise expectation (see Figure \ref{FigPoisson}); to determine this number we take the average
of bin sizes between 10 and 80 binned points. In general, we have noticed that 
the RMS$\times$$\beta$ of the residuals is a superior metric to determine the best aperture size/reference star combination
than simply the RMS; for our near-infrared photometry, which generally suffers from high sky background compared
to the optical, the RMS of the residuals generally reaches a minimum for
relatively small apertures, as one is able to reduce the impact of the high sky background. However, these small
apertures often suffer from time-correlated red-noise (high $\beta$s), as during moments of poor seeing or tracking errors,
a small fraction of the light falls outside these small apertures.
% Usually the RMS$\times$$\beta$ does the best job of identifying the appropriatebalancing the need to reduce systematic errors from light
% falling outside the aperture,
A small complication for our CFHT/WIRCam Ks-band photometry of KIC 1255, however,
is that, as noted above, we believe that some of the correlated red-noise we observe is genuine,
as it reproduces, in part, the short-term variations
visible in the 
{\it Kepler} optical photometry.
Therefore we qualitatively noticed that the most useful metrics were
the RMS of the residuals for the first CFHT/WIRCam transit, and
the RMS of the residuals multiplied by $\beta$ for the second CFHT/WIRCam transit; this combination
produced the most satisfactory results. For this reason,
we used these two different metrics to determine the best reference star ensemble
and aperture size combination below.

In Figure \ref{FigKICFidelity} we present the correlations with aperture radius and the number of stars in our 
reference star ensemble for both our first and second CFHT transit of KIC 1255b. 
For both cases we display the variation in RMS, and the variation in RMS multiplied by $\beta$,
for a variety of aperture radii and different number of reference stars in the ensemble,
as well as the associated variation in the associated KIC 1255b transit depths, as parameterized by $A$.
In the top set of panels we display the RMS, RMS$\times$$\beta$, and $A$ values for a variety of aperture
sizes for the best 7-star reference 
ensemble\footnote{We use a 7-star reference ensemble
as this gave the minimum RMS of the residuals for the first transit, and the minimum 
RMS$\times$$\beta$ of the residuals for the second transit.}.
In the bottom set of panels we display the 
RMS, RMS$\times$$\beta$, and $A$ values for ensembles of different numbers of reference stars,
for a 7.0-pixel aperture (first CFHT transit) and a 9.0-pixel aperture (second CFHT transit).

We note that the differences in the parameter of interest $A$ are modest for most combinations
of reference star ensembles, and aperture radii. Nevertheless, it is important to scale up our errors
on $A$ to take into account these correlations. To determine the transit depth of KIC 1255b
once correlations with reference star ensemble and aperture radius are taken into account, $A_{Corr}$,
we take the weighted mean of all values with an RMS no greater than 10\% above the minimum RMS for our first
transit\footnote{We therefore average aperture radii of 7.0-9.5 pixels, and 4-11 reference
stars for our first CFHT transit.},
and for our second CFHT transit we take all values with an RMS$\times$$\beta$ no greater than
40\%\footnote{
We use 40\%, rather than the 10\% utilized for the other reference star/aperture size combinations, due to the
fact that the aperture radii 5.5 and 6.0 display such small RMS$\times$$\beta$ (the top right panel
of Figure \ref{FigKICFidelity}). This is due to the fact that for these aperture radii
a different 7-star reference ensemble was chosen automatically by the routine to have the smallest RMS. This
reference star ensemble features stars that are not as bright, and display considerably worse RMS$\times$$\beta$
values for larger aperture sizes. We prefer the 9.0-pixel reference star ensemble, and therefore
present it in Figure \ref{FigCFHT}.
}
and 10\% above
the minimum RMS$\times$$\beta$ value (these are denoted by the dotted horizontal line in Figure \ref{FigKICFidelity})
for the various aperture radii, and number of reference stars, respectively\footnote{We therefore average aperture
radii of 5.5, 6.0 and 7.5-12.5 pixels, and 4-11 reference stars.}. 
To determine the error on $A_{Corr}$ we calculate the mean error of all 
the values used to determine $A_{Corr}$, and add to this value, in quadrature, the standard deviation of 
the $A$ values. The $A_{Corr}$ values for the first and second CFHT transit
are given in Table \ref{TableParams}. The error on $A_{Corr}$ for both transits
has increased marginally, compared to that on $A$ before the correlations with the number of stars
in the reference ensemble, and the aperture size, were taken into account.

%EMULATEAPJCHANGE
\begin{deluxetable*}{ccccccc}
%\begin{deluxetable}{ccccccc}
\setlength{\tabcolsep}{0.01in} 
\tablecaption{CFHT, {\it HST}, \& {\it Kepler} photometry of KIC 1255}
\tabletypesize{\scriptsize}
\tablehead{
\colhead{Parameter} 	& \colhead{CFHT} 		& \colhead{{\it Kepler}}	& \colhead{CFHT}	& \colhead{Kepler}	& \colhead{{\it HST}}		& \colhead{Kepler}	\\	
\colhead{}		& \colhead{Transit \#1}		& \colhead{Transit \#1}		& \colhead{Transit \#2}	& \colhead{Transit \#2}	& \colhead{Transit}	& \colhead{Transit \#3}	\\
}
\startdata
$A$					& \ValueZerocfhtkeplerCFHTOne$^{+\ValueZeroPluscfhtkeplerCFHTOne}_{-\ValueZeroMinuscfhtkeplerCFHTOne}$ 		& \ValueZerocfhtkeplerKeplerOne$^{+\ValueZeroPluscfhtkeplerKeplerOne}_{-\ValueZeroMinuscfhtkeplerKeplerOne}$ 	& \ValueZerocfhtkeplerCFHTTwo$^{+\ValueZeroPluscfhtkeplerCFHTTwo}_{-\ValueZeroMinuscfhtkeplerCFHTTwo}$ 		& \ValueZerocfhtkeplerKeplerTwo$^{+\ValueZeroPluscfhtkeplerKeplerTwo}_{-\ValueZeroMinuscfhtkeplerKeplerTwo}$ 	& \ValueZerohstkeplerHST$^{+\ValueZeroPlushstkeplerHST}_{-\ValueZeroMinushstkeplerHST}$		& \ValueZerohstkeplerKepler$^{+\ValueZeroPlushstkeplerKepler}_{-\ValueZeroMinushstkeplerKepler}$	\\
$A_{Corr}$				& \ValueZeroCFHTAlphaI$\pm$\ValueZeroErrorCFHTAlphaI 								& n/a 														& \ValueZeroCFHTAlphaII$\pm$\ValueZeroErrorCFHTAlphaII								& n/a														& n/a												& n/a													\\
$t_{transit}$ (BJD-2456150)			& \ValueTwocfhtkeplerCFHTOne$^{+\ValueTwoPluscfhtkeplerCFHTOne}_{-\ValueTwoMinuscfhtkeplerCFHTOne}$ 		& \ValueTwocfhtkeplerKeplerOne$^{+\ValueTwoPluscfhtkeplerKeplerOne}_{-\ValueTwoMinuscfhtkeplerKeplerOne}$ 	& \ValueTwocfhtkeplerCFHTTwo$^{+\ValueTwoPluscfhtkeplerCFHTTwo}_{-\ValueTwoMinuscfhtkeplerCFHTTwo}$ 		& \ValueTwocfhtkeplerKeplerTwo$^{+\ValueTwoPluscfhtkeplerKeplerTwo}_{-\ValueTwoMinuscfhtkeplerKeplerTwo}$	& \ValueTwohstkeplerHST\tablenotemark{a}							& \ValueTwohstkeplerKepler\tablenotemark{a}								\\
$y_o$						& \ValueThreecfhtkeplerCFHTOne$^{+\ValueThreePluscfhtkeplerCFHTOne}_{-\ValueThreeMinuscfhtkeplerCFHTOne}$ 	& \ValueThreecfhtkeplerKeplerOne$^{+\ValueThreePluscfhtkeplerKeplerOne}_{-\ValueThreeMinuscfhtkeplerKeplerOne}$ & \ValueThreecfhtkeplerCFHTTwo$^{+\ValueThreePluscfhtkeplerCFHTTwo}_{-\ValueThreeMinuscfhtkeplerCFHTTwo}$ 	& \ValueThreecfhtkeplerKeplerTwo$^{+\ValueThreePluscfhtkeplerKeplerTwo}_{-\ValueThreeMinuscfhtkeplerKeplerTwo}$	& \ValueThreehstkeplerHST$^{+\ValueThreePlushstkeplerHST}_{-\ValueThreeMinushstkeplerHST}$	& \ValueThreehstkeplerKepler$^{+\ValueThreePlushstkeplerKepler}_{-\ValueThreeMinushstkeplerKepler}$	\\
3$\sigma$ upper-limit on $A$		& \ValueZeroThreeSigmaPluscfhtkeplerCFHTOne									& \ValueZeroThreeSigmaPluscfhtkeplerKeplerOne									& \ValueZeroThreeSigmaPluscfhtkeplerCFHTTwo									& \ValueZeroThreeSigmaPluscfhtkeplerKeplerTwo									& \ValueZeroThreeSigmaPlushstkeplerHST								& \ValueZeroThreeSigmaPlushstkeplerKepler								\\
\enddata
% \ValueTwohstkeplerKepler$^{+\ValueTwoPlushstkeplerKepler}_{-\ValueTwoMinushstkeplerKepler}$
\tablenotetext{a}{We fix $t_{transit}$ to the predicted mid-point of the transit for this analysis, due to the fact we are unable to detect the transit on this occasion.}
\label{TableParams}
%EMULATEAPJCHANGE
%\end{deluxetable}
\end{deluxetable*}

\subsection{{\it HST} photometry and imaging}

 On 2013 February 6 (UTC) we observed a transit of KIC 1255 with the 
{\it Hubble Space Telescope} ({\it HST}) Wide Field Camera 3 (WFC3; \citealt{Dressel10})
over 5 orbits ({\it HST} Proposal \#GO-12987, P.I. = S. Rappaport).
The first orbit was devoted to high angular resolution 
imaging observations of KIC 1255 in order
to rule-out nearby background and foreground objects or companions to this object; these observations
are detailed in $\S$\ref{SecHSTImaging}. The second
through fifth {\it HST} orbits were devoted to F140W ($\lambda$$\sim$1.39 $\mu m$)
photometry of the transit of KIC 1255; these observations
are detailed below in $\S$\ref{SecHSTPhotometry}.

We use the calibrated, flat-fielded {\it flt} files from WFC3's calwf3 reduction pipeline. %INSERTHERE isn't there a reference?
%Should I say what the flt files do
% bad pixel flagging, reference pixel subtraction, zero-read subtraction, dark current subtraction, non-linearity correction, 
%flat-field correction, as well as gain and photometric calibration.

\subsubsection{{\it HST} photometry}
\label{SecHSTPhotometry}

\begin{figure}
\centering
\includegraphics[scale=0.45, angle = 90]{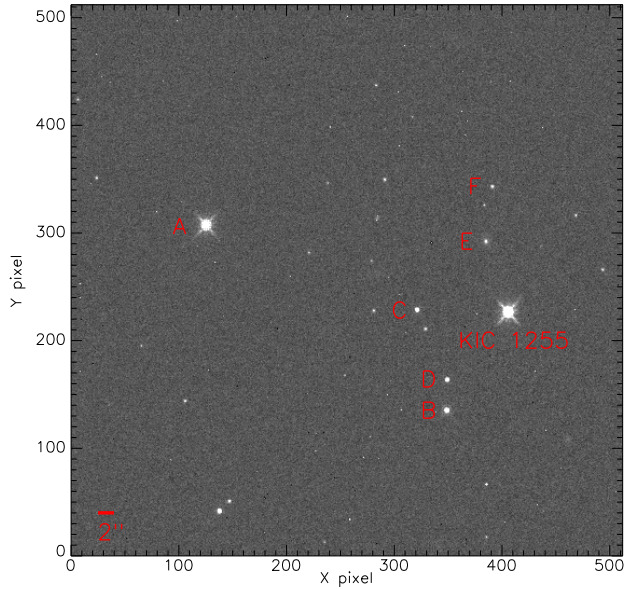}
\caption[KICRaelert]
	{
	The full frame array of our {\it HST}/WFC3 1.4 $\mu m$ photometry. The target
	star, as well as the handful of nearby reference stars (as listed in Table \ref{TableRefStars})
	that we perform photometry on to rule out false-positive scenarios,
	are labeled with letters from A-F.%STARRAELERTCHANGE 
	}
\label{FigEntireFrame}
\end{figure}

\begin{figure}
\centering
\includegraphics[scale=0.60, angle = 270]{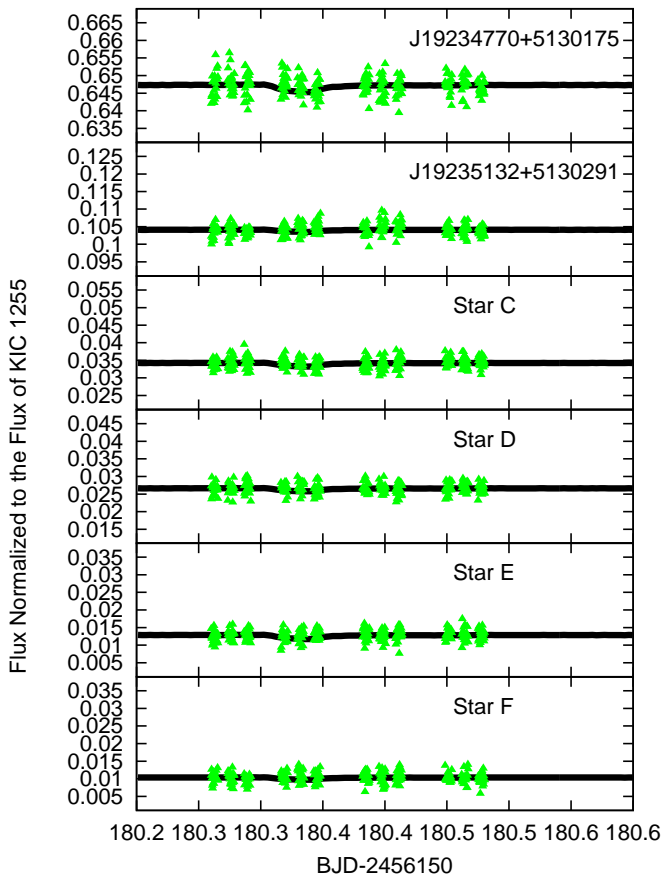}
\caption[KICHSTRefStars]
	{
	{\it HST} F140W filter photometry of various reference stars (listed in Table \ref{TableRefStars}) to KIC 1255.
	The solid black line denotes the 3$\sigma$ upper-limit on the transit depth
	obtained by scaling the mean of the short-cadence, phase-folded {\it Kepler} light curve (see $\S$\ref{SecKepler}).
	}

\label{FigHSTRefStarsPhotometry}
\end{figure}

%STARRAELERTCHANGE
%EMULATEAPJCHANGE
\begin{deluxetable*}{ccccccc}
%\begin{deluxetable}{cccccc}
\tablecaption{{\it HST} photometry of reference stars to KIC 1255}
\tablehead{
\colhead{Star Name} 			& \colhead{Angular Separation} 		& \colhead{$A$}		& \colhead{$t_{transit}$\tablenotemark{a}}	& \colhead{$y_o$}	& \colhead{3$\sigma$ upper-limit}	& \colhead{Percentage Brightness}\\	
\colhead{} 				& \colhead{From KIC 1255 (\arcsec)} 	& \colhead{}		& \colhead{(BJD-2456150)}			& \colhead{}		& \colhead{ on $A$}			& \colhead{of KIC 1255 at $\sim$1.39 $\mu m$}\\	
}
\startdata
Star A (2Mass J19234770+5130175)	& 39.15					& \ValueZerorefstarA$^{+\ValueZeroMinusrefstarA}_{-\ValueZeroPlusrefstarA}$			& \ValueTworefstarA				& \ValueThreerefstarA$^{+\ValueThreePlusrefstarA}_{-\ValueThreeMinusrefstarA}$			& \ValueZeroThreeSigmaPlusrefstarA	& 64.7	\\
Star B (2Mass J19235132+5130291)	& 13.40					& \ValueZerorefstarB$^{+\ValueZeroMinusrefstarB}_{-\ValueZeroPlusrefstarB}$			& \ValueTworefstarB				& \ValueThreerefstarB$^{+\ValueThreePlusrefstarB}_{-\ValueThreeMinusrefstarB}$			& \ValueZeroThreeSigmaPlusrefstarB	& 10.4	\\
Star C					& 11.34					& \ValueZerorefstarC$^{+\ValueZeroMinusrefstarC}_{-\ValueZeroPlusrefstarC}$			& \ValueTworefstarC				& \ValueThreerefstarC$^{+\ValueThreePlusrefstarC}_{-\ValueThreeMinusrefstarC}$			& \ValueZeroThreeSigmaPlusrefstarC	& 3.4	\\
Star D					& 10.69					& \ValueZerorefstarD$^{+\ValueZeroMinusrefstarD}_{-\ValueZeroPlusrefstarD}$			& \ValueTworefstarD				& \ValueThreerefstarD$^{+\ValueThreePlusrefstarD}_{-\ValueThreeMinusrefstarD}$			& \ValueZeroThreeSigmaPlusrefstarD	& 2.7	\\
Star E					& 8.39					& \ValueZerorefstarE$^{+\ValueZeroMinusrefstarE}_{-\ValueZeroPlusrefstarE}$			& \ValueTworefstarE				& \ValueThreerefstarE$^{+\ValueThreePlusrefstarE}_{-\ValueThreeMinusrefstarE}$			& \ValueZeroThreeSigmaPlusrefstarE	& 1.3	\\
Star F					& 14.28					& \ValueZerorefstarF$^{+\ValueZeroMinusrefstarF}_{-\ValueZeroPlusrefstarF}$			& \ValueTworefstarF				& \ValueThreerefstarF$^{+\ValueThreePlusrefstarF}_{-\ValueThreeMinusrefstarF}$			& \ValueZeroThreeSigmaPlusrefstarF	& 1.0	\\
% Star G					& 10.48					& \ValueZerorefstarF$^{+\ValueZeroMinusrefstarF}_{-\ValueZeroPlusrefstarF}$			& \ValueTworefstarF				& \ValueThreerefstarF$^{+\ValueThreePlusrefstarF}_{-\ValueThreeMinusrefstarF}$			& \ValueZeroThreeSigmaPlusrefstarF		\\
%\ValueTworefstarA$^{+\ValueTwoPlusrefstarA}_{-\ValueTwoMinusrefstarA}$ - this would be if I fit the t_transit
\enddata
\tablenotetext{a}{We fix $t_{transit}$ to the predicted mid-point of the transit for this analysis.}
\label{TableRefStars}
%EMULATEAPJCHANGE
\end{deluxetable*}
%\end{deluxetable}

 We obtained {\it HST}/WFC3 photometry over four {\it HST} orbits 
of the transit of KIC 1255 in the F140W filter ($\lambda$$\sim$1.39 $\mu m$)
on 2013 February 6 (UTC).
The exposure times were 4.27 seconds; observations were obtained every 22.98 seconds, resulting in a duty cycle of
$\sim$19\%.
252 observations were obtained over the four {\it HST} orbits, or 63 observations per orbit.
We performed aperture photometry, as described above in $\S$\ref{SecCFHT}.
We use an aperture radius of 5.75 pixels;
we do not subtract the background with an annulus, or otherwise.
% , and to subtract the background we use
% an annulus with an inner and outer radius of 25 and 33 pixels.
The results are nearly identical whether
we do or do not subtract the background with an annulus with our aperture photometry.
The {\it HST} photometry of KIC 1255 is presented in Figure \ref{FigHSTPhotometry},
as well as the best-fit scaled version of the mean short-cadence, phase-folded {\it Kepler} photometry ($g(t)$; $\S$\ref{SecKepler});
the associated parameters are listed in Table \ref{TableParams}.
We are unable to detect the transit of KIC 1255 in either our
{\it HST} photometry or the simultaneous {\it Kepler} photometry (on 2013 February 6). We place
a 3$\sigma$ upper limit on the transit depths on these two occasions of:
$A$ $<$ \ValueZeroThreeSigmaPlushstkeplerHST \ 
for our {\it HST} photometry, and $A$ $<$ \ValueZeroThreeSigmaPlushstkeplerKepler \ for our {\it Kepler} Transit \#3.
We note that the transit of KIC 1255b that we observed with {\it HST} and {\it Kepler} happened to be during a
stretch of time where the KIC 1255b transit depth was below detectability for a number of transits
in a row (approximately $\sim$5 days before and after our observed transit).

 To rule out whether the transit that we associate with KIC 1255 is actually due to a nearby object,
we also perform photometry on all relatively bright stars nearby to KIC 1255 in the {\it HST} photometry. 
By using the centroiding analysis methods of \citep{Jenkins10}
on the {\it Kepler} photometry of KIC 1255, \citet{Rappaport12} report 
that any background source that is causing the transits that we
associate with KIC 1255 must be within 2\arcsec \ of KIC 1255.
As we discuss in $\S$\ref{SecHSTImaging} we rule out sufficiently bright (at least 1\% of the flux of KIC 1255)
reference stars down to 0.2\arcsec \ from KIC 1255, and a cursory inspection of Figure \ref{FigEntireFrame}
indicates there are no sufficiently bright reference stars out to 11\arcsec \ from KIC 1255.
Objects fainter than 1\% of KIC 1255, if fully occulted, would not be able to account for the greater than 1\% transit
depth observed in the {\it Kepler} photometry. 
Despite the fact that during our {\it HST} observations our {\it HST} and {\it Kepler} photometry did not 
display a detectable transit of KIC 1255b, we 
nevertheless perform photometry on all relatively bright stars that are within 20\arcsec of KIC 1255,
and a few select stars of comparable brightness to KIC 1255 that are captured
in our {\it HST} photometry.
F140W photometry on all these stars from orbits 2-5 
is presented in Figure \ref{FigHSTRefStarsPhotometry}.
We display these stars in Figure \ref{FigEntireFrame}, which shows the median full-frame
image of all our {\it HST} F140W photometry.
The 3$\sigma$ upper-limits on the transit depth for these stars
are given in Table \ref{TableRefStars}. None
of these stars displays obvious behaviour that would 
suggest that they serve as a false positive for the characteristic photometry that
we associate with
KIC 1255b\footnote{Not that we would necessarily expect them to
show such behaviour during our {\it HST} observations, as the 
KIC 1255b transit was undetectable during these observations.}. 

% We use the below F775W imaging observations, which are the closest in wavelength to that of the {\it Kepler} bandpass,
% to find all nearby stars to KIC 1255 that are at least $\sim$1\% the brightness of KIC 1255.

\subsubsection{{\it HST} high-angular resolution imaging}
\label{SecHSTImaging}

%EMULATEAPJCHANGE
\begin{figure*}
%\begin{figure}
\centering
%EMULATEAPJCHANGE, 0.90 for aastex, 1.00 for emulateApJ
\includegraphics[scale=1.00, angle = 90]{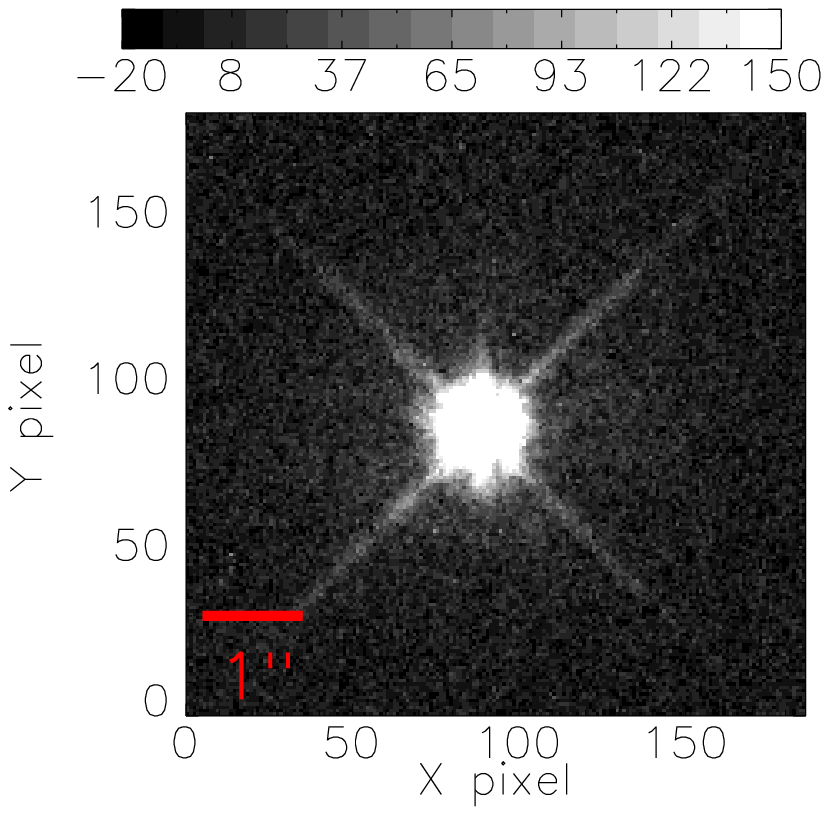}
\includegraphics[scale=1.00, angle = 90]{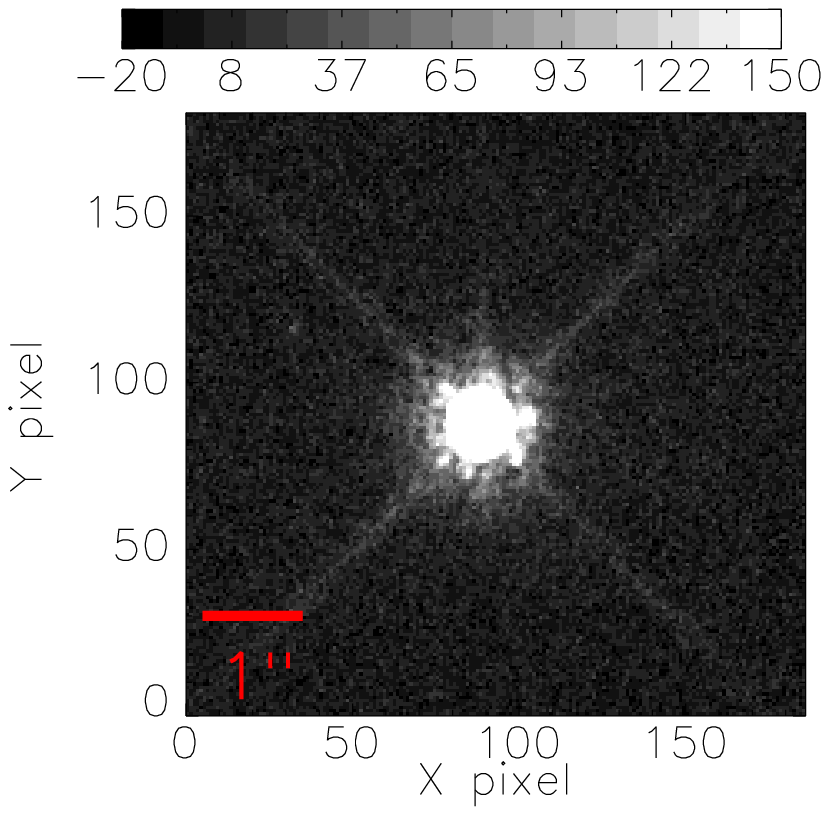}
\includegraphics[scale=1.00, angle = 90]{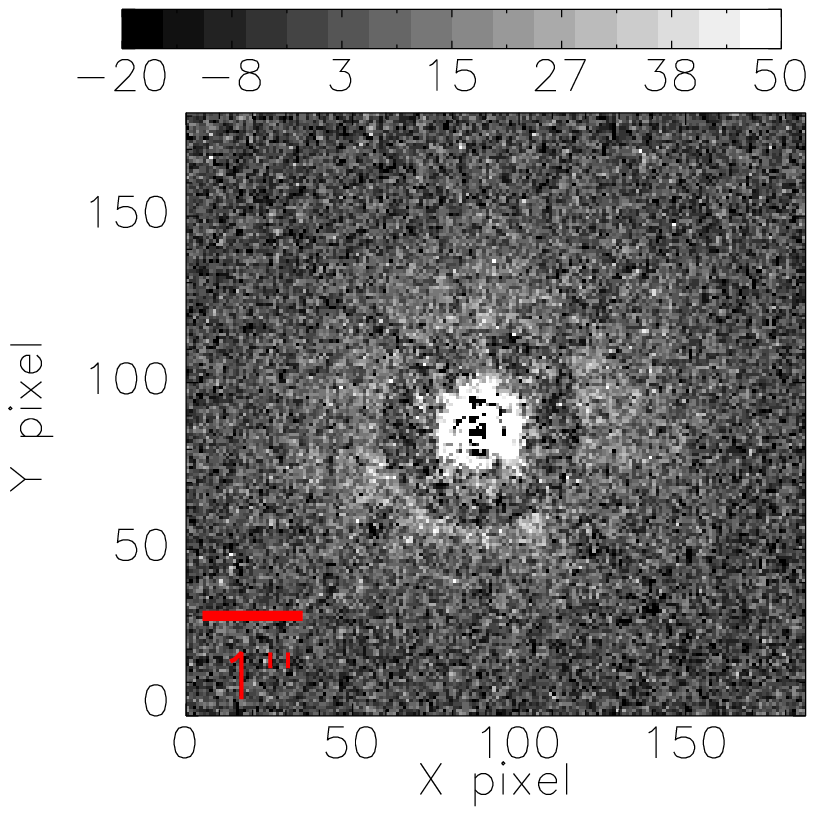}
\includegraphics[scale=1.00, angle = 90]{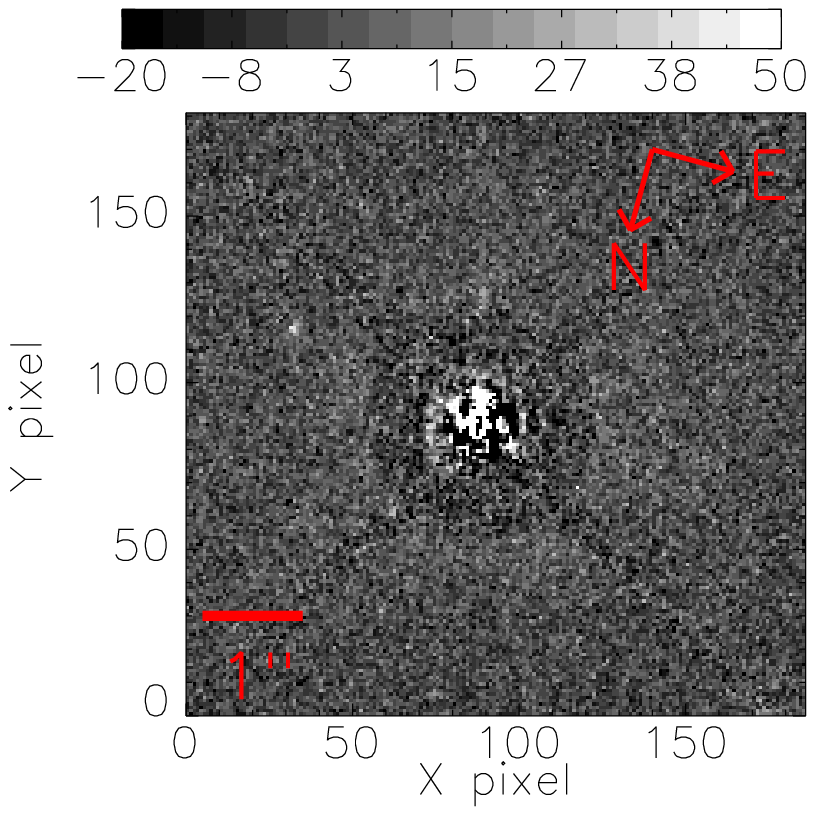}
\caption[KICHSTImaging]
	{
		{\it HST}/WFC3 high angular resolution imaging of KIC 1255 in the F555W channel (left),
		and the F775W channel (right). The top row of images are the AstroDrizzled images,
		while the bottom row of images are the PSF subtracted images.
		For both set of images, the scale is indicated at top, in units of electrons.
		The compass rose for all panels is given in the lower-right panel.
		For reference in the F775W difference image (bottom-right)
		the object at 10 o'clock at 2{\farcs}2 separation 
		is at a delta-magnitude of 9.1, and is thus too faint, by a factor of more than $\sim$50, to cause the transit 
		we associate with KIC 1255b.
	}
\label{FigHSTImaging}
%EMULATEAPJCHANGE
\end{figure*}
%\end{figure}

\begin{figure}
\centering
\includegraphics[scale=0.95, angle = 270]{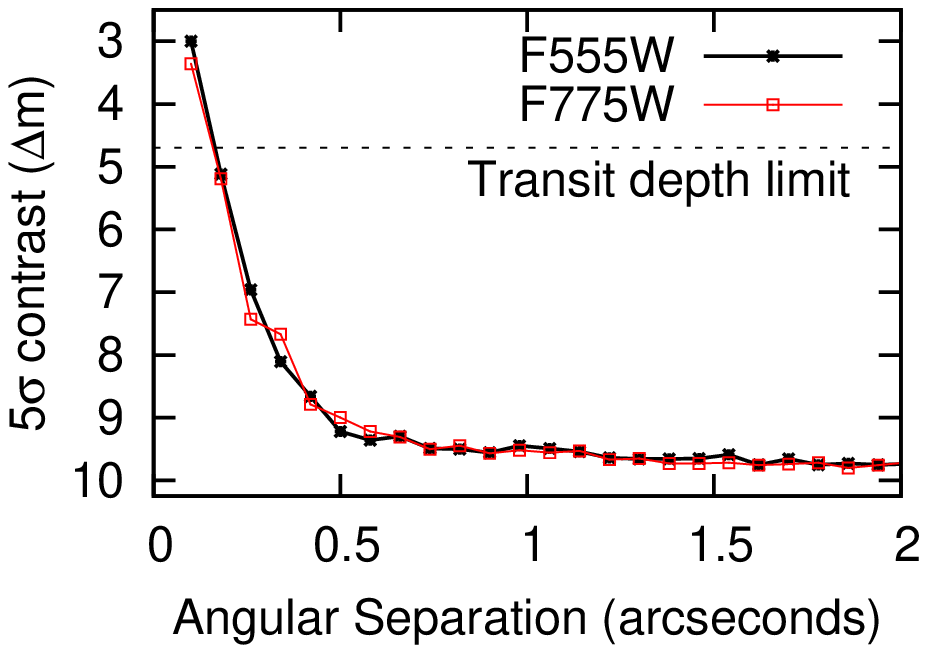}
\caption[KICHSTImaging]
	{
		Contrast limits for our {\it HST}/WFC3 high angular resolution imaging
		on nearby background/foreground companions to KIC 1255
		in difference of magnitude in the 
		F555W channel (red solid line, unfilled squares),
		and the F775W channel (black solid line, crosses).
		The horizontal dashed line
		denotes the limiting magnitude of the faintest object
		that could produce the transits we associated with KIC 1255b.
		We rule out companions
		bright enough to cause the transit we associate with KIC 1255b for separations
		greater than 0.2\arcsec (5$\sigma$).
	}
\label{FigHSTImagingSensitivity}
\end{figure}

{\it HST} high-angular resolution imaging observations to search for nearby companions
to KIC 1255 were obtained with the WFC3 instrument
in the following filters in the Infrared channel (IR):
F125W ($\lambda$$\sim$1.25 $\mu m$),
F140W ($\lambda$$\sim$1.39 $\mu m$),
F160W ($\lambda$$\sim$1.54 $\mu m$),
and the following filters in the ultraviolet and visible (UVIS) channel:
F555W ($\lambda$$\sim$0.531 $\mu m$),
F775W ($\lambda$$\sim$0.765 $\mu m$).
We present results of the reduction 
and analysis of the F555W and F775W channels here. %, via analysis led by R.L.G. and K.M.S.

  Through program GO-12893, which imaged some of {\em Kepler's}
best Kepler Objects of Interest (KOIs) in terms of small planet candidates on long orbits,
we had available F555W and F775W observations of many targets
taken with exactly the same dither pattern and essentially the same signal-to-noise.
We searched for a subset of the GO-12893 observations for which
the target: (a) seemed to be an isolated single star, (b) had
a $g - r$ color similar to KIC 1255, and (c) for which the
{\em HST} focus offset matched that for KIC 1255.  Visits number
60 (KIC 8150320), 94 (KIC 4139816), and 98 (KIC 5942949)
met these criteria and were
processed with AstroDrizzle \citep{Fruchter2010} to the same
0{\farcs}0333 scale used for the KIC 1255 images.  All observations
consisted of four dithered exposures in which the target was
kept just under detector saturation, and a fifth exposure at
twice saturation to bring up the wings.  In each case the
drizzle combination was done to provide a final image well
centered on a pixel.  The KIC 1255 direct imaging results are shown
in the upper panels of Figure \ref{FigHSTImaging} for 460 seconds in F555W and
330 seconds in F775W.  The combined images had a FWHM of
$\sim$0{\farcs}075, and are given in units of electrons\footnote{The
negative electron values observable in Figure \ref{FigHSTImaging}, are likely due to the noise in the dark
and sky frames that are subtracted to produce the frames seen here.}.  The observations in the three GO-12893
controls were averaged together to define a PSF for each
filter, then scaled to the intensity of the KIC 1255 images
and subtracted to provide the difference images shown in the
lower panels of Figure \ref{FigHSTImaging}, in units of electrons.
The subtraction is performed to a radius of 1{\farcs}0
at F555W and 1{\farcs}2 at F775W; this subtraction is extended to $\sim$3\arcsec
along the diagonal diffraction spikes.  For reference, the
object at 10 o'clock at 2{\farcs}2 separation in Figure 8
of the F775W difference image is at a delta-magnitude
of 9.1.

  The difference images were used to derive 5$\sigma$
detection limits as a function of offset distance by evaluating
what fluctuations in 3$\times$3 0{\farcs}0333 pixels would
stand out at this level relative to the scatter in successive
0{\farcs}08 annuli starting with one centered at 0{\farcs}1
\citep{Gilliland2011}.
These contrast limits are presented in Figure \ref{FigHSTImagingSensitivity}.
The HST imaging ruled out potential sources of background false positives down to
0{\farcs}2
for a delta-magnitude of 4.7 (equivalent to the maximum observed Kepler transit depth),
thereby reducing the original
2\arcsec \ radius area 
in which these background/foreground candidates could exist by 99\% (i.e., $<$0.2\arcsec).

%Comment on why some of the values are negative. Could include.
% I think there are two factors coming into play leading to pixels at negative, and therefore nominally non-physical
% values in the direct image.  The first is that these are sky subtracted, one doesn't have to go out very far before
% the rms on the sky plus readout noise exceeds the physical level from the PSF wings -- leading to minor negative values.
% A second factor will come from subtraction of available dark frames (scaled by relative exposure times), the
% darks are never perfectly contemporaneous with the data, and I probably didn't make efforts to redo the reduction
% later with the best possible darks, that will lead to a small (but noticeable) subset of pixels having over-subtracted
% dark values.  I don't think there is any issue of consequence with this.

\subsection{Keck Adaptive Optics Imaging}

%EMULATEAPJCHANGE
\begin{figure*}
%\begin{figure}
\centering
%EMULATEAPJCHANGE, 0.55 and 0.45 for aastex, 0.62 and 0.52 for emulateApJ
\includegraphics[scale=0.62, angle = 0]{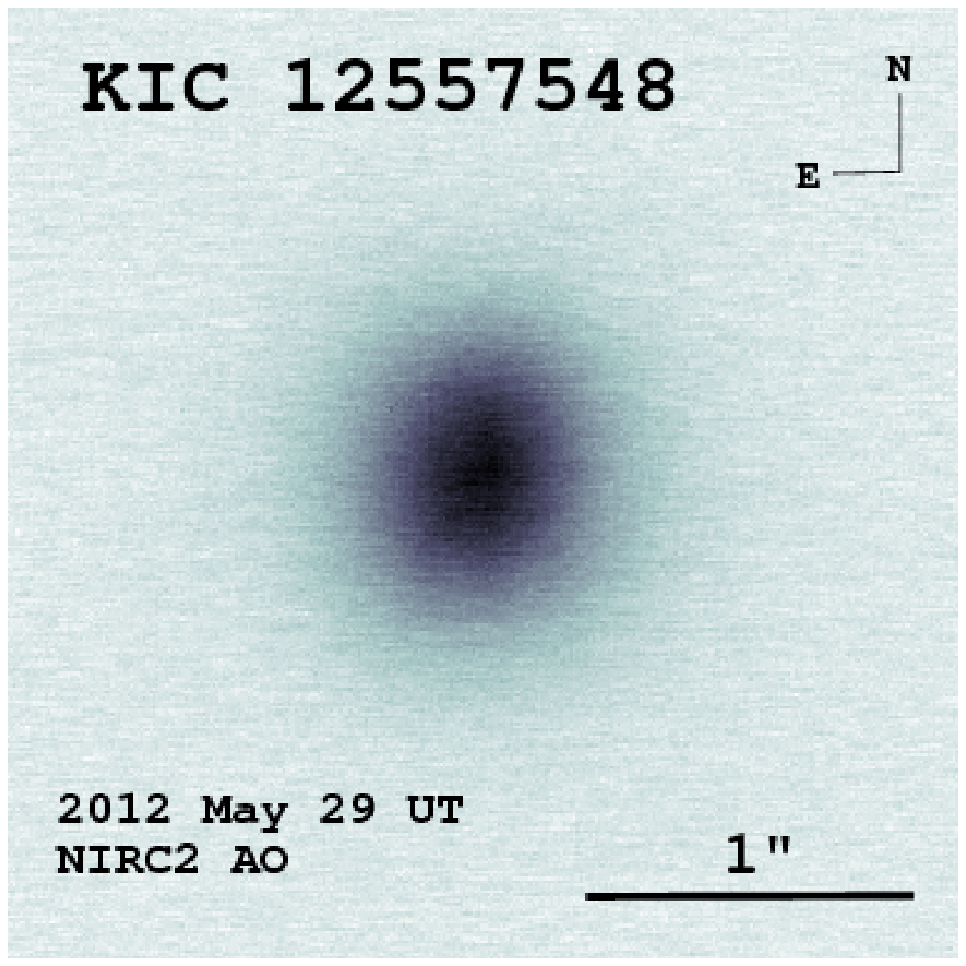}
\includegraphics[scale=0.52, angle = 0]{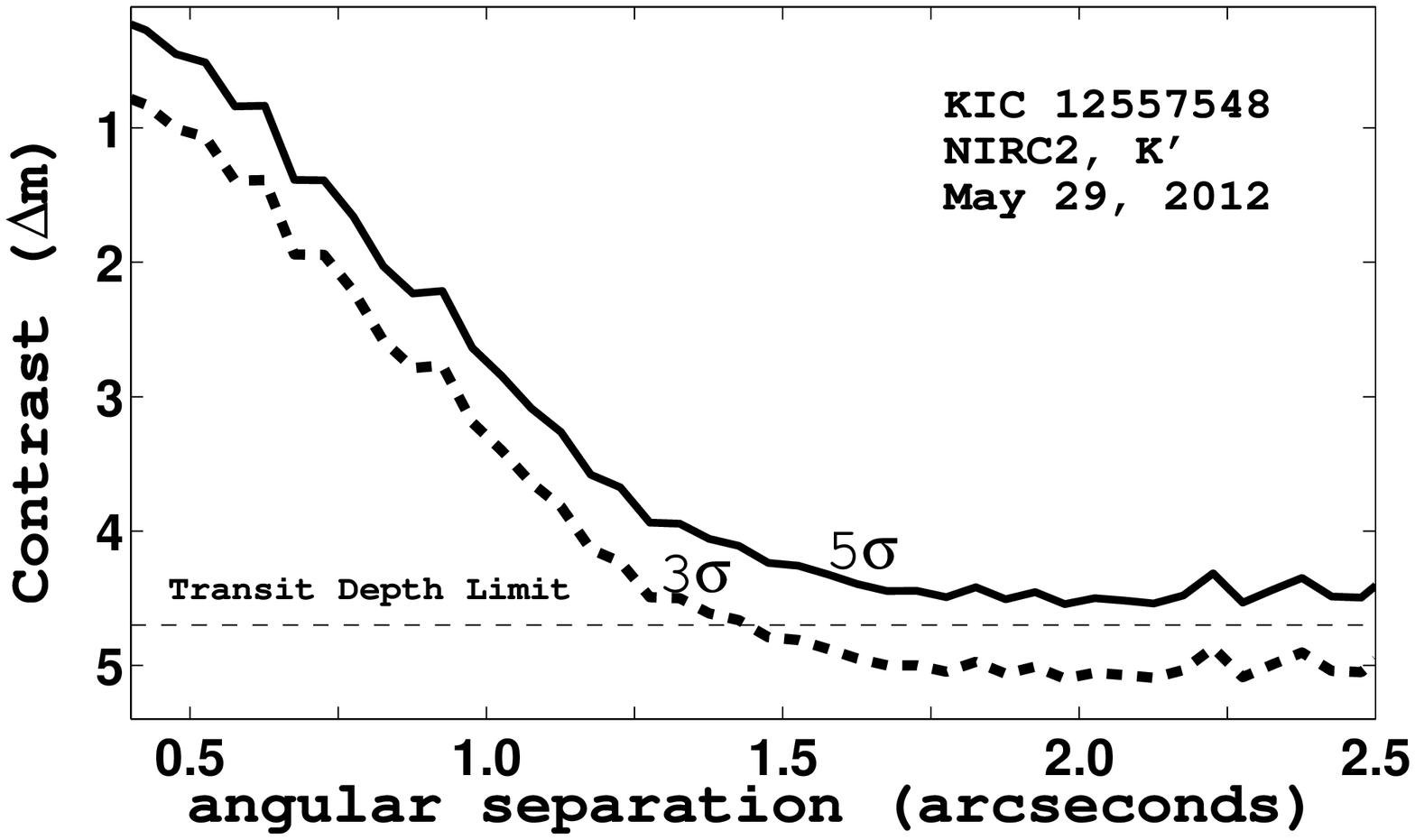}
\caption[KICAOCrepp]
	{
	Keck/NIRC2 high angular resolution imaging (left)
	of KIC 1255 in the K' filter ($\lambda_c=$2.124 $\mu m$). 
	The associated contrast limits on nearby background/foreground companions
	to KIC 1255 (right) in difference of magnitude in the K' filter are given with 5$\sigma$ (solid line) and
	3$\sigma$ confidence (thick-dashed line); the thin-dashed horizontal line
	denotes the limiting magnitude of the faintest object that could produce the transits we associated with KIC 1255b.
	We rule out companions that bright at this wavelength for separations
	greater than 1.4\arcsec (3$\sigma$).

	}
\label{FigAOCrepp}
%EMULATEAPJCHANGE
\end{figure*}
%\end{figure}

%INSERTHERE by J.C.

We obtained high angular-resolution, near-infrared images of KIC 1255 on 2012 May 29 UT using NIRC2 (PI: Keith Matthews) and the Keck II adaptive
optics (AO) system (Wizinowich et al. 2000). Observations were acquired in natural guide star mode with the K' filter
($\lambda\sim$2.124 $\mu m$). Given the faintness of KIC 1255 (R=15.30), we opened the deformable mirror loops and
applied tip/tilt correction commands only. Images were recorded using the narrow camera mode which provides a 10 mas
plate scale. A standard 3-point dither pattern was executed to remove background radiation from the sky and instrument
optics. A total on-source integration time of 360 seconds was obtained from 6 separate frames. Images were processed
using standard techniques to replace hot pixel values, flat-field, subtract the background, and align and coadd frames
\citep{Crepp2012}.

Figure \ref{FigAOCrepp} shows the final reduced image (left panel)
along with sensitivity to off-axis sources (right panel). No
obvious companions were noticed in either raw or processed frames. Comparing residual scattered light levels to the
stellar peak intensity, our NIRC2 observations rule out the presence of possible photometric contaminants, at
differential flux values comparable to and brighter than the maximum Kepler transit
depth\footnote{See $\S$\ref{SecLimits} for an explanation of why this optical limit ($\Delta m=4.7$)
is likely valid in the near-infrared.}
($\Delta m=4.7$),
for separations $>1.4$\arcsec \ at $3\sigma$.

\subsection{Keck Radial Velocities}
\label{SecRVs}

\begin{figure}
\includegraphics[scale=0.45, angle = 270]{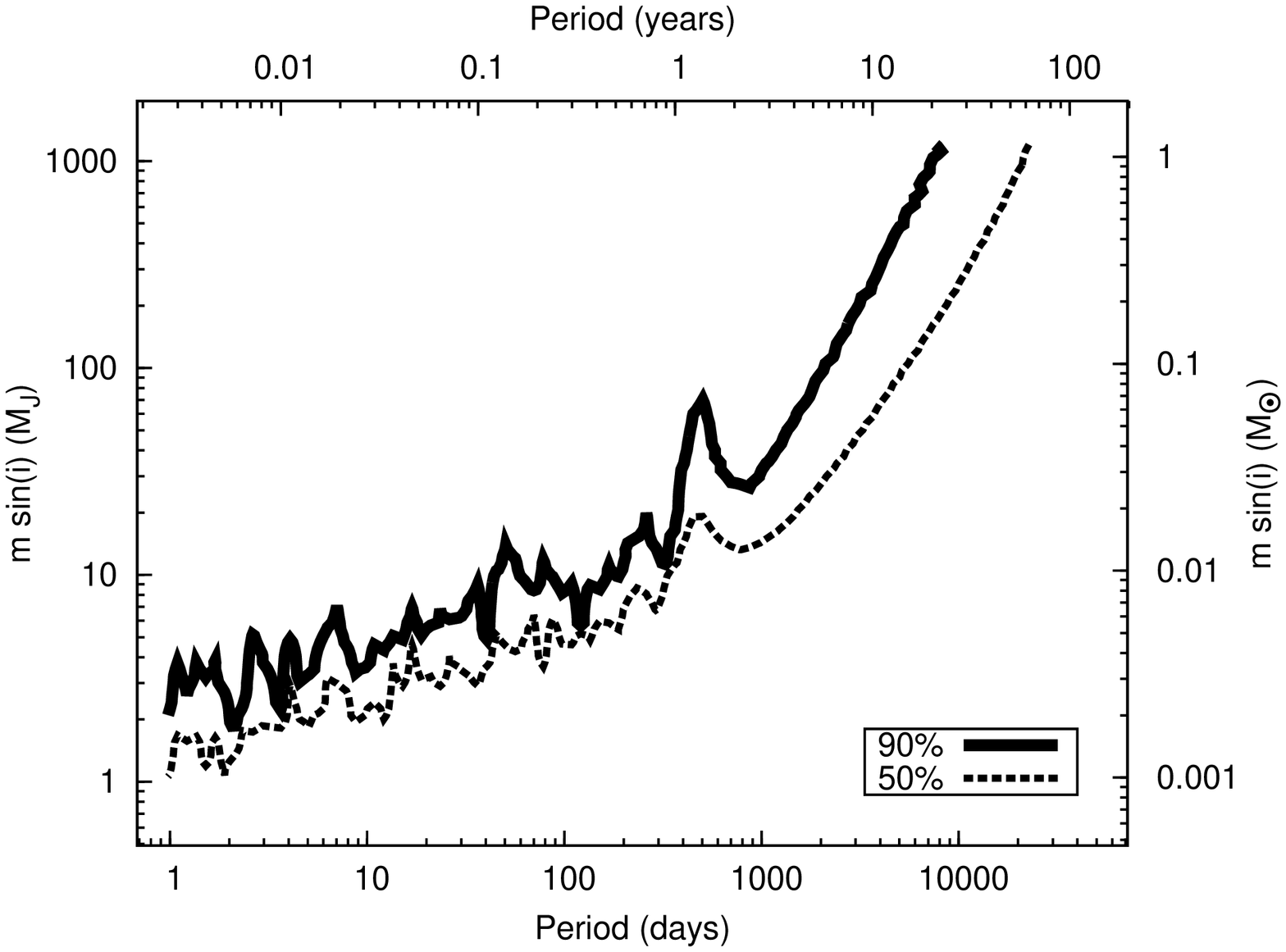}
	\caption{
	Constraints on the minimum mass, $m \, \sin$$i$, and period, $P$, of
	possible companions to KIC 1255.  These are based
	on the Keck/HIRES RV measurements that we obtained with telluric-calibrated
	spectra. Artificial radial velocity signals were injected into the
	data and were recovered 50\% of the time (dashed line), and 90\%
	of the time (solid-line) with strong confidence (at a threshold greater
	than 3$\sigma$). We are able to rule-out giant planetary mass companions at
	periods $\lesssim 40$ days, and low-mass stellar companions ($m \, \sin i \sim 0.2
	\, M_\odot$) for $\lesssim 10$ year orbits with 90\% confidence.
	}
\label{FigRVLimits}
\end{figure}

We obtained five high-resolution spectra of KIC 1255 using HIRES
\citep{Vogt94} on the Keck I Telescope to
measure absolute radial velocities.  
The exposures were 5--10 minutes in duration and achieved signal-to-noise ratios of 
12--16 per pixel in $R$-band (on blaze).
We followed the standard techniques of the California Planet Survey 
for the reduction and sky subtraction of spectra \citep{Batalha11}.  
We measured absolute radial velocities (RVs) with the telluric oxygen A and B bands 
(759.4--762.1 nm and 686.7--688.4 nm) as a wavelength reference using the method of \citet{Chubak12}.  
The photon-weighted times of observation, RVs, and errors are listed in Table \ref{TableRVs}.
Individual measurements carry 0.1 km~s$^{-1}$ uncertainties, 
as demonstrated by bright standard star measurements \citep{Chubak12} as well 
as measurements of faint stars such as Kepler-78 ($V$ = 12; \citealt{Howard13}).

The RV measurements of KIC 1255 span 512 days and have an RMS of 0.17 km~s$^{-1}$.  
We searched for accelerations in the RV time series that could indicate a stellar-mass, long-period 
companion.  A linear least-squares fit to the data yields a statistically insignificant slope 
of $-0.16 \pm 0.90$ km~s$^{-1}$~yr$^{-1}$.  
We then injected artificial RV signals for a hypothetical companion, with a given 
period, $P$, and minimum mass, $m \, \sin$$i$ (where $i$ is the orbital inclination of
the companion), into the data and determined with what confidence we would be able to detect these companions.
Zero-eccentricity orbits were assumed, and $4000$ trials were drawn with a random orbital phase.
Inspired by a similar method used by \citet{Bean10},
signals were judged to be detected if the $\chi^2$ of the RV fit after subtracting the mean, $\chi_{RV}^2$,
were greater than the $\chi^2$ of a straight line fit,
$\chi_{straight}^2$\footnote{$\chi_{straight}^2$=11.25 with 4 degrees of freedom.},
plus an amount corresponding to 3$\sigma$ confidence 
for four degrees of freedom (from our five data-points) -- 
that is if: $\chi_{RV}^2$ $>$ $\chi_{straight}^2$ + 16.3. 
The resulting 50\% and 90\% confidence limits 
are given in Figure \ref{FigRVLimits}; stronger confidence limits (e.g. 99.73\%)
result in considerably higher-masses, as the time-gaps in the RV data allow for even very
massive companions to slip through a small fraction of the time. 
Also, allowing eccentric orbits would result in less constraining (higher mass) limits at a given orbital period.
Our analysis rules out giant planetary-mass companions ($m \, \sin$$i$ $\sim$ 13 M$_J$ [Jupiter masses])
for $\lesssim 40$ day orbits, 
and low-mass stellar companions
($m \, \sin$$i$ $\sim$ 0.2 M$_\odot$ or greater) for $\lesssim 10$ year orbits with 90\% 
confidence\footnote{We note that if there was an additional companion in the KIC 1255 system (other than the candidate planet),
the light-time effect of the host star orbiting
around the barycentre of the system \citep{Montalto10}, would induce periodic variability in the 
transit-timing of the candidate planet. For long period companions this might manifest itself
as long-term variations in the period of the candidate planet;
such changes may have been detectable in the {\it Kepler} data, given the stringent limits on the
lack of change in the orbital period
reported by \citet{Budaj13} and \citet{vanWerkhoven13}.}.
We note an M-dwarf companion
with a mass less than 0.2 M$_\odot$
would verge on the limit of being too faint to cause the 1\% transit depth we associate with KIC 1255.

\subsubsection{Radial Velocity Limit on KIC 1255b's minimum mass}

\begin{figure}
\includegraphics[scale=0.45, angle = 270]{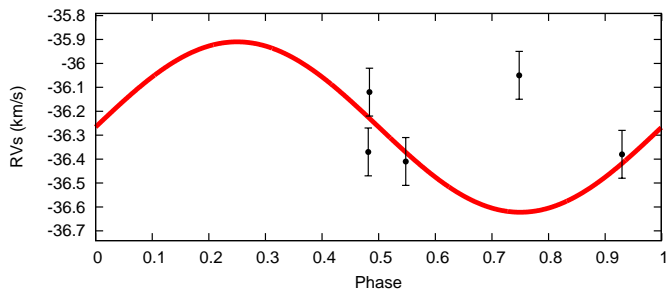}
	\caption{
	The Keck/HIRES RV measurements phased to the orbital period of the candidate planet
	($P$$\sim$0.6536 $d$). The red-solid line represents
	the radial velocity fit corresponding to the upper-limit on the mass
	of the candidate planet ($m \, \sin i$ $\lesssim$ 1.2 M$_J$) %PLANETMASSLIMIT
	that we can rule-out with 3$\sigma$ certainty, for a circular orbit.
	}
\label{FigPhaseRVs}
\end{figure}

% CHANGEHERE - remember you have to take into effect the duration of the RV observations. Yes we do this, already.

Our Keck RV measurements also allow us to attempt to detect, or set an upper limit on,
the mass of the candidate planet KIC 1255b. Our Keck RVs phased to the orbital period of KIC 1255b are displayed in Figure 
\ref{FigPhaseRVs}.
Similar to the techniques used above, we then insert 
circular (zero eccentricity) RV signals at KIC 1255b's orbital period, and phase, until these signals 
exceed the 
chi-squared limit discussed in $\S$\ref{SecRVs}.
Our 3$\sigma$ upper limit on
KIC 1255b's minimum mass from the Keck RVs is then approximately 1.2 Jupiter-masses %PLANETMASSLIMIT
($m \, \sin i$ $\lesssim$ 1.2 M$_J$), %PLANETMASSLIMIT
as displayed in Figure \ref{FigPhaseRVs}.
We note that given the sparse RV sampling, allowing eccentric orbits would result in a much higher
upper limit on KIC 1255b's minimum mass. 
Given KIC 1255b's close orbit with its parent star, 
one would not necessarily expect KIC 1255b to have an eccentric orbit; however
the planet's high mass loss rate, suggests a short life-time for the planet in its present
orbit \citep{PerezBeckerChiang13},
leaving open the possibility that the planet's orbit may not have circularized yet.

\begin{deluxetable}{ccc}
\tabletypesize{\footnotesize}
\tablecaption{Telluric-calibrated Radial Velocities from Keck-HIRES}
\tablewidth{0pt}
\tablehead{ 
    \colhead{JD -- 2440000}		& \colhead{Absolute Radial Velocity}		& \colhead{Uncertainty}  \\
    \colhead{}				& \colhead{(km s$^{-1}$)}			& \colhead{(km s$^{-1}$)} 
}
\startdata
16020.05982  &  $-$36.05  &  0.10  \\
16028.02128  &  $-$36.38  &  0.10  \\
16076.09278   &  $-$36.12  &  0.10  \\
16110.07628   &  $-$36.37  &  0.10  \\
16532.97013   &  $-$36.41  &  0.10  \\
\enddata
\label{TableRVs}
\end{deluxetable}

\section{Multiwavelength Photometric Results}
\label{SecPhotometricAnalysis}

\begin{deluxetable}{cc}
\tablecaption{$A$ ratios for KIC 1255}
\tabletypesize{\scriptsize}
\tablehead{
\colhead{Data \& Transit \#} 		& \colhead{ Ratio} \\	
}
\startdata
CFHT \& {\it Kepler} Transit \#1		& $A_{Corr}$/$A_{Kepler}$ = \RatioCFHTCorrelatedI \ $\pm$ \RatioErrorCFHTCorrelatedI		\\
CFHT \& {\it Kepler} Transit \#2		& $A_{Corr}$/$A_{Kepler}$ = \RatioCFHTCorrelatedII \ $\pm$ \RatioErrorCFHTCorrelatedII					\\
CFHT \& {\it Kepler} Transits Combined		& $A_{CFHT}$/$A_{Kepler}$ = \RatioBOTHCFHTKepler \ $\pm$ \RatioBOTHErrorCFHTKepler		\\
{\it HST} \& {\it Kepler} Transit		& $A_{HST}$/$A_{Kepler}$ = \Ratiohstkepler \ $\pm$ \RatioErrorhstkepler 			\\
\enddata
\label{TableAlpha}
\end{deluxetable}

 Our simultaneous {\it Kepler} ($\sim$0.6 $\mu m$) and CFHT ($\sim$2.15 $\mu m$) photometry,
and our simultaneous {\it Kepler} and {\it HST} ($\sim$1.4 $\mu m$) photometry
allow us to compare the transit depths from the optical to the near-infrared.
The ratio of the {\it Kepler} to the CFHT transit depth is
similar in both our first \& second CFHT and {\it Kepler} observations
($A_{Corr}$/$A_{Kepler}$ = \RatioCFHTCorrelatedII \ $\pm$ \RatioErrorCFHTCorrelatedI \
on 2012 August 13 2012 and 
$A_{Corr}$/$A_{Kepler}$ = \RatioCFHTCorrelatedI \ $\pm$ \RatioErrorCFHTCorrelatedI \
on 2012 September 1 2012).
The weighted mean of the $A$ ratio
from both observations is:
$A_{CFHT}$/$A_{Kepler}$ = \RatioBOTHCFHTKepler \ $\pm$ \RatioBOTHErrorCFHTKepler.
For the simultaneous {\it HST} and {\it Kepler} observation we are only 
able to return a null-detection of the transit depth at those wavelengths; the associated
ratio of the {\it HST} to {\it Kepler} transit depths is: $A_{HST}$/$A_{Kepler}$ = \Ratiohstkepler \ $\pm$ \RatioErrorhstkepler.
Therefore, we
can only say that there is no evidence for strongly different transit depths
at these wavelengths.
We summarize the ratios at these wavelengths in Table \ref{TableAlpha}.

\section{Limits on nearby companions to KIC 1255b and false positive scenarios}
\label{SecLimits}

 Although we know of no viable binary or higher-order multiple scenario
that could explain the unusual photometry we observe for KIC 1255, we note that our high-angular resolution imaging, 
RVs, and multiwavelength photometry place strict limits on companions to KIC 1255 and thus on suggested 
false positive scenarios.

We searched for nearby companions to KIC 1255 with our {\it HST}/WFC3 and Keck/NIRC2 high-angular resolution imaging.
With a maximum transit depth of 1.3\% of the stellar flux, the maximum magnitude differential between a background object
and KIC 1255 that could be causing the behaviour we observe is 4.7 magnitudes.
Therefore our {\it HST}/WFC3 high angular resolution F555W,
and F775W ($\lambda$$\sim$0.765 $\mu m$)
imaging\footnote{The F555W ($\lambda$$\sim$0.531 $\mu m$) and F775W ($\lambda$$\sim$0.765 $\mu m$) bracket the $\sim$0.6 $\mu m$
midpoint of the {\it Kepler} bandpass.}
is able to rule out companions this bright with 5$\sigma$ confidence down to 0.2\arcsec \ angular separation from
KIC 1255b (see Figure \ref{FigHSTImagingSensitivity}).

In the near-infrared, given the CFHT to {\it Kepler} transit depth ratio we measure
here ($A_{CFHT}$/$A_{Kepler}$ = \RatioBOTHCFHTKepler \ $\pm$ \RatioBOTHErrorCFHTKepler), we can expect that any object
must be no more than
4.7 magnitudes fainter than KIC 1255b at these wavelengths as well; therefore for our Keck/NIRC2 imaging
we are able to rule out companions this bright down to 1.4\arcsec \ separation at 3$\sigma$ confidence.

Our radial-velocity observations allow us to rule-out low mass stellar companions 
($\sim$ 0.2 M$_\odot$) for reasonably edge-on orbits, 
for periods less than $10$ years (this corresponds to $\lesssim$4 AU using the 0.7 $M_{\odot}$ stellar
mass reported by \citealt{Rappaport12}). Our high angular-resolution {\it HST} imaging limit of 0.2\arcsec, corresponds to 
$>$94 AU at the $\sim$470 parsec distance of KIC 1255 quoted by \citet{Rappaport12}.
We therefore note there is little viable parameter space
(only companion separations, $s$, of 4 AU$\lesssim$ $s$ $\lesssim$ 94 AU
remain viable) for a binary or higher-order multiple companion
to KIC 1255b that could
be masquerading as a false-positive for the photometry of KIC 1255 that we associate with a disintegrating low-mass planet.

 Lastly, we note that the fact that our near-infrared and optical photometry report similar transit depths, also allows
us to place an additional constraint on 
hierarchical triple or background binary configurations with stars of different spectral types. 
For instance, if the photometry we associate with KIC 1255b was somehow due to a 
pair of late M-dwarfs (an effective temperature of $T_{\rm eff}$$\sim$3000 K)
eclipsing one another,
whose light was diluted by the K-dwarf star KIC 1255 ($T_{\rm eff}$ $\sim$ 4400 K; \citealt{Rappaport12}),
the $\sim$1\% optical transit depths, would result in $\sim$6\% transit depths at 2.15 $\mu m$ - a depth we can rule with very
high confidence ($\sim$25$\sigma$). %CHANGEHERECONFIRMIFNUMBERCHANGES
Similar limits could be set on a background binary that is of a different spectral type from KIC 1255.
A background binary of similar spectral type to KIC 1255 remains a possible false positive; however it is unclear
how such a scenario would explain the variable transit depths, and asymmetric transit profile we observe with this candidate planet.

 With these stringent limits on false positive scenarios, we therefore conclude 
that the disintegrating low-mass planet scenario is the simplest explanation suggested to date for the {\it Kepler} photometry, 
and our
multiwavelength photometry, RVs, and high angular-resolution imaging.

\section{Discussion}

\subsection{Size of grains in the dust tail of KIC 1255b}

The wavelength dependence of extinction by dust grains can provide information on their size and, in some cases, 
their composition \citep[e.g., for interstellar grains see][]{Mathis1977}.
This is due to the fact that the efficiency of scattering generally diminishes 
as the observational wavelength approaches the
approximate particle circumference \citep{HansenTravis74},
For this reason, the nearly identical transit depths we measure at 
$\lambda$ $\sim$0.6 $\mu m$ with {\it Kepler} and 
at $\lambda$ $\sim$2.15 $\mu m$ with CFHT/WIRCam allow us to set a lower-limit on the largest particles 
in the hypothetical dust tail trailing KIC 1255b. We set this lower-limit on the size
assuming that all the dust particles are a single, identical size in $\S$\ref{SecSingleSizeGrains},
and for a distribution of particle sizes in $\S$\ref{SecGrainSizeDistributions}.

%  Following a brief discussion of the spectral dependence of extinction by dust grains,
% we define a measure of the sensitivity of extinction to wavelength, and apply it first to individual grains and consider
% the implications for the measurements presented earlier for KIC1255b. We then expand the discussion to distributions of grains
% of varying size. 

%
\begin{figure}[h]
\includegraphics[width=0.50\textwidth,angle=0]{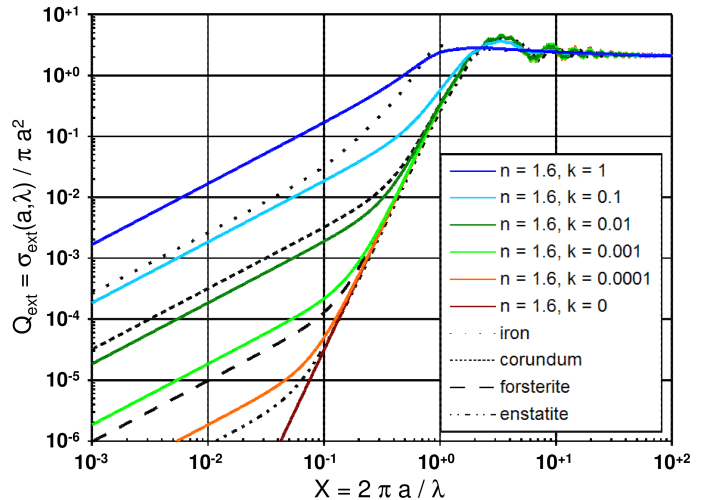}\\
	\caption{The extinction efficiency, $Q_{ext}$ (Mie extinction cross sections, normalized to the geometric area of the grain), as a function of the nondimensional
	grain size parameter, $X = 2 \pi a / \lambda$, where $a$ is the radius of the grain, and $\lambda$ is the wavelength of observation.
	The dotted, short-dashed, long-dashed, and dot-dashed lines indicate the extinction efficiency of
	iron, corundum, forsterite
	and enstatite, as calculated as described in the text.}
\label{FigMieExtinctionPlot}
\end{figure}

\subsubsection{Spectral Dependence of Extinction}
\label{SecSpectralDependenceOfExtinction}

\begin{deluxetable}{ccc}
\setlength{\tabcolsep}{0.01in} 
\tablecaption{Wavelength Extinction and Sensitivity Limits}
\tabletypesize{\scriptsize}
\tablehead{
\colhead{Parameter region} 		& \colhead{$\sigma_{ext}(a,\lambda)$} 		& \colhead{$\alpha(a,\lambda_1,\lambda_2)$}	\\	
}
\startdata
$a \gg \lambda / 2 \pi$ 		& $2 \pi a^2$ 					& $0$					\\
$a \ll \lambda / 2 \pi$ and $k = 0$ 	& $\propto$ $a^6 / \lambda^4$ 			& $4$        				\\
$a \ll \lambda / 2 \pi$ and $k \ne 0$	& $\propto$ $a^3 / \lambda$ 			& $1$    				\\
\enddata
\label{TableSensitivityLimits}
\end{deluxetable}

The extinction (scattering plus absorption) of light by dust grains is a function of the wavelength, $\lambda$, of the light,
the size of the grains, denoted by the grain radius, $a$, and the complex index of refraction of the dust material,
where $n$ denotes the real component and $k$ the imaginary component.
While there is also some dependence of extinction upon the shape of the dust grains,
for simplicity we will only consider spherical grains here, and employ the 
Mie algorithm presented by \cite{Bohren+Huffman1983} to calculate their cross sections.
% Fig.\,\ref{FigMieExtinctionPlot} plots the
% extinction cross section (normalized by the geometric cross section),
% $\sigma_{ext}(a,\lambda) / \pi a^2$, as a
% function of the nondimensional size, $X \equiv  2 \pi a / \lambda$, 
% for grains with $n = 1.6$, and $k = 0$ (no absorption) and $k$ varying from
% $0.0001$ (low absorption) to $1$ (highly absorbing).  
Fig.\,\ref{FigMieExtinctionPlot} plots the extinction efficiency (the ratio of the extinction
cross section to the geometric cross section),  
$Q_{ext}(a,\lambda) = \sigma_{ext}(a,\lambda) / (\pi \, a^2)$, as a function of the
nondimensional size,  $X = 2 \pi a / \lambda$, for grains with n =
1.6, and k = 0 (no absorption) and k varying from 0.0001 (low
absorption) to 1 (highly absorbing).

We note that for large grains ($X \gg 1$; the right-hand side of the plot),
the extinction cross-section is approximately constant
% ($\sigma_{ext}(a,\lambda) \rightarrow 2 \pi a^2$)
($Q_{ext}(a,\lambda) = \sigma_{ext}(a,\lambda)/ (\pi \, a^2) \rightarrow 2$)
and only slightly dependent on wavelength,
regardless of the absorption, $k$. For wavelengths longer or approaching the approximate particle circumference
(the left side of the plot), the amount of extinction
nominally depends on the amount of absorption \citep[e.g.,][]{vandeHulst1981}.
Nonetheless, even for a great deal of absorption (i.e., $k$=1; the dark blue line in Figure \ref{FigMieExtinctionPlot}), the
extinction cross-section falls sharply for wavelengths longer than the approximate 
particle circumference\footnote{For $k \ne 0$, the extinction can be approximated by $\sigma_{ext}(a,\lambda) / \pi a^2 \propto k \, X$.};
for low levels of absorption or none at all (the light green, 
orange, and red lines in Figure \ref{FigMieExtinctionPlot}),
the fall-off is even
steeper\footnote{For $ k \equiv 0$ the extinction can be approximated by $\sigma_{ext}(a,\lambda) / \pi a^2 \propto X^4$,
a relation found by \cite{Rayleigh1871}. For $k \ne 0$ there may be a transition region between the two
extremes where $\sigma_{ext}(a,\lambda) / \pi a^2 \propto X^4$.}.
The asymptotic limits apparent in the plot are listed in Table \ref{TableSensitivityLimits}.
Therefore we surmise based on 
the nearly equal {\it Kepler} and CFHT transit depths, and thus the near equal levels of extinction between these two 
wavelengths, that our observations probe the right-hand side of the plot; we can therefore set a lower-limit on
the largest particles in the hypothetical dust-tail trailing KIC 1255b.

For the scattering calculations that follow,
we assume typical values for the complex index of refraction, $n$ and $k$, 
based on the values for typical Earth-abundant refractory materials, such as olivines and pyroxenes.
% The main advantage of olivines and pyroxenes is that they can endure the
% relatively high temperatures that are expected in the putative tail 
% trailing KIC 1255b.
Across our wavelength range of interest ($\lambda$=0.6 to 2.15 $\mu m$),
for these materials, the imaginary component of the index of refraction is typically small,
$k$ $\lesssim$ 0.02,
while the real component of the index of refraction 
is often approximately n$\sim$1.6 (\citealt{Kimura02} and references therein).
We also repeat our scattering calculations for 
a number of materials that have previously
been suggested to be responsible for the dust supposedly trailing
KIC 1255b. Four such materials, suggested by \citet{Budaj13},
are  
forsterite (Mg$_{2}$SiO$_{4}$; a silicate from the olivine family),
enstatite (MgSiO$_3$; a pyroxene without iron),
pure iron, and corundum (Al$_2$O$_3$; a crystalline form of aluminium oxide).
Three of these materials have similar complex indices of refraction
across our
wavelength range of interest ($\lambda$=0.6 to 2.15 $\mu m$):
that is n$\sim$1.6 and $k$$<$10$^{-4}$ for enstatite \citep{Dorschner95},
n$\sim$1.6 and $k$$<$10$^{-3}$ for forsterite \citep{Jager03},
n$\sim$1.6 and $k$$<$0.04 for corundum \citep{Koike95}.
Pure iron is an outlier
with n$\sim$2.9 - 3.9 and $k$$\sim$3.4 - 7.0 \citep{Ordal88}
for wavelengths from $\lambda$=0.6 to 2.15 $\mu m$.
In Figure \ref{FigMieExtinctionPlot} we also display the resulting extinction
efficiency, $Q_{ext}(a,\lambda)$, of these
materials\footnote{The curves presented in Figure \ref{FigMieExtinctionPlot} represent averages 
of the extinction efficiency, $Q_{ext}$,
for wavelengths of 0.6, 1.0, 1.6, 2.0 and 2.15 $\mu
m$ (except for iron, which omits the 0.6 $\mu m$ calculation because it is below the tabulated
index of refraction values),
using index of refraction
values for enstatite from \citet{Dorschner95}, forsterite from \citet{Jager03}, corundum from \citet{Koike95},
and iron from \citet{Ordal88}.}.

\subsubsection{Extinction Wavelength Dependence}
\label{SecExtinctionWavelengthDependence}

To parameterize the amount of extinction between our two wavelengths, $\lambda_1$ and $\lambda_2$, we
employ the \r{A}ngstr\"{o}m exponent, $\alpha(a,\lambda_1,\lambda_2)$, a measure of the dependence of extinction on wavelength\footnote{Defined by \cite{Angstrom1929} in the context of dust in the Earth's atmosphere.}, defined as follows:

\begin{equation}
\alpha(a,\lambda_1,\lambda_2) \equiv - \frac{ \log{ [ \sigma_{ext}(a,\lambda_2) / \sigma_{ext}(a,\lambda_1) ] } }{ \log{ ( \lambda_2 / \lambda_1 ) } }
\label{eqn:WavelengthDependence01}
\end{equation}

The ratio of the transit depths in Table \ref{TableAlpha} is approximately the ratio of the extinctions at these two wavelengths,
$\sigma_{ext}(a,\lambda_2) / \sigma_{ext}(a,\lambda_1) = \RatioBOTHCFHTKepler \pm \RatioBOTHErrorCFHTKepler$.
Therefore, the associated value of
$\alpha(a,0.6 \, {\rm \mu m}, 2.15 \, {\rm \mu m}) = \SValue$ with
ranges $\SValuePlusOneSigma$ to $\SValueMinusOneSigma$ ($1 \sigma$),
$\SValuePlusTwoSigma$ to $\SValueMinusTwoSigma$ ($2 \sigma$),
and $\SValuePlusThreeSigma$ to $\SValueMinusThreeSigma$ ($3 \sigma$).

% CHANGE_HERE FIX
% The ratio of the transit depths in Table \ref{TableAlpha} is simply the ratio of the extinctions at these two wavelengths,
% $\sigma_{ext}(a,\lambda_2) / \sigma_{ext}(a,\lambda_1) = \RatioBOTHCFHTKepler \pm \RatioBOTHErrorCFHTKepler$.
% Therefore, the associated value of
% $S(a,0.6 \, {\rm \mu m}, 2.15 \, {\rm \mu m}) =$ \SValue \ $\pm$ \SValueOneSigma \ (1$\sigma$).
% %, or $S(a,0.6 \, {\rm \mu m}, 2.15 \, {\rm \mu m}) =$ \SValue$\pm$\SValueTwoSigma \ (2$\sigma$).
% $\alpha(a,0.6 \, {\rm \mu m}, 2.15 \, {\rm \mu m}) = \SValue$ with ranges $\SValuePlusOneSigma$ to $\SValueMinusOneSigma$ ($1 \sigma$), $\SValuePlusTwoSigma$ to $\SValueMinusTwoSigma$ ($2 \sigma$), and $\SValuePlusThreeSigma$ to $\SValueMinusThreeSigma$ ($3 \sigma$).

\subsubsection{Single Size Grains}
\label{SecSingleSizeGrains}

\begin{figure}[h]
\includegraphics[width=0.48\textwidth,angle=0]{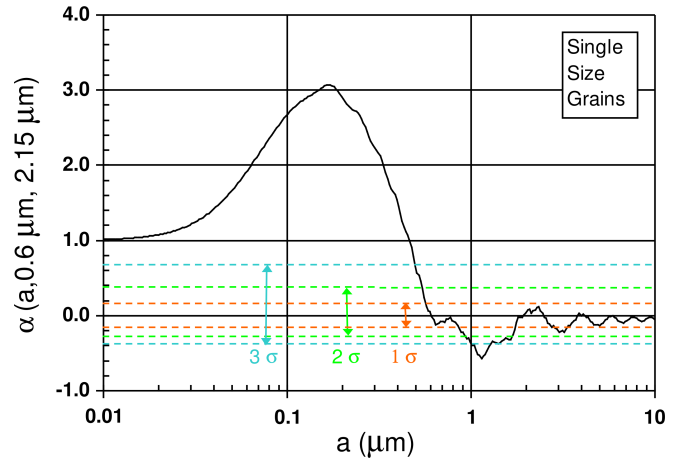}\\
	\caption{
		Plot of the \r{A}ngstr\"{o}m exponent for spherical grains of radius, $a$. We assume an index of refraction of $n = 1.6$, and 
		an imaginary component of the index of refraction of  $k = 0.02$.
		The horizontal orange, green, and blue dashed lines show the 1$\sigma$, 2$\sigma$, and 3$\sigma$ limits
		on $\alpha(a,\lambda_1,\lambda_2)$, respectively,
		from $\S$\ref{SecExtinctionWavelengthDependence}.  Grains of a single-size have to be larger
		than $a$$\gtrsim \ParticleSizeCare$ $\mu m$ (3$\sigma$) to be consistent with
		the ratio of our CFHT/WIRCam and {\it Kepler} transit depths.
	}
\label{FigSingleParticleAngstromPlot}
\end{figure}

What is the maximum size of particles in the hypothetical
dust tail of KIC 1255b if the tail is composed solely of single-size spherical particles?
Fig. \ref{FigSingleParticleAngstromPlot} plots Mie calculations of $\alpha(a,0.6 \, {\rm \mu m}, 2.15 \, {\rm \mu m})$ for grains
with $n = 1.6$, $k = 0.02$, and radius $a$.
The 1, 2, and 3$\sigma$ error bars on $\alpha(a,0.6 \, {\rm \mu m}, 2.15 \, {\rm \mu m})$
are displayed with the dashed orange, green, and blue lines, respectively, in Figure \ref{FigSingleParticleAngstromPlot}.
Therefore, for a dust tail consisting of single size grains their radius would have to be
$\gtrsim \ParticleSizeCare$ ${\rm \mu m}$ (3$\sigma$).
 
We have also reproduced these scattering
calculations for four materials that have been suggested to make-up
the particles trailing KIC 1255b (these are corundum, pure iron, forsterite and enstatite),
rather than for our hypothetical
$n=1.6$ and $k$=0.02 material. Due to the close agreement between
the complex index of refraction of 
forsterite, enstatite and corundum
and our assumed $n=1.6$ and $k$=0.02 values, the size 
limit on single-sized 
forsterite, enstatite or corundum particles is indistinguishable from our hypothetical 
material; that is such grains
must be $\gtrsim \ParticleSizeCare$ ${\rm \mu m}$ (3$\sigma$ confidence).
Pure iron, on the other hand,
which has a complex index of refraction that differs significantly from the above values,
results in a less stringent limit on single size iron particles; pure iron particles 
would have to be $\gtrsim 0.2$ ${\rm \mu m}$ (3$\sigma$). However,
due to iron's high vapour
pressure\footnote{Iron
has a vapor pressure $\sim$50 times greater than that for olivines \citep{PerezBeckerChiang13}. 
The survival of olivines have already been called into question at the extreme $\sim$2000 $K$
temperatures expected in the tail trailing KIC 1255b \citep{Rappaport12}.},
we find it doubtful that pure iron particles could
survive the high temperatures of a dust tail trailing KIC 1255b in the first place
without sublimating.
We therefore quote only our $\gtrsim \ParticleSizeCare$ ${\rm \mu m}$ (3$\sigma$)
size-limit on single-size particles,
henceforth\footnote{Obviously, if the grains in the putative tail trailing KIC 1255b are
composed of a material with very different optical properties than what we have
assumed -- $n$$\sim$1.6 with a small $k$ ($\lesssim$0.1) -- this could result in a different minimum size
than we have presented here. We are unaware of a material that is likely to be present trailing KIC 1255b,
with very different optical properties
than what we have assumed,
and that is likely to survive the high temperatures in the tail trailing KIC 1255b without sublimating.}.

%According to Kimura et al. (2002) the mass loss due to sublimation is proportional to the vapor pressure. Therefore higher vapor pressures at a given temperature result in higher mass loss rates due to sublimation.

%Note sand particles start at 60 microns, up to 2 mm.

% footnote{The odd behaviour with particle size that we observe in Figure \ref{FigSingleParticleSensitivityPlot} can be explained by
% substituting the asymptotic forms for $\sigma_{ext}(a,\lambda)$ into this equation produces the sensitivity
% results shown in the third column of Table \ref{tab:SensitivityLimits} for the three regions discussed above.
% In the limit that $a \gg \max(\lambda_1,\lambda_2) / 2 \pi$ ($\approx 0.4$ in the plot), $S(a,\lambda_1,\lambda_2) \rightarrow 0$.
% Oscillations in $\sigma_{ext}(a,\lambda)$ extends the range of $S(a,\lambda_1,\lambda_2)$ somewhat above $0$. 
% Since $k \ne 0$, for $a \ll \min(\lambda_1,\lambda_2) / 2 \pi$ ($\approx 0.01$ in the plot), $S(a,\lambda_1,\lambda_2) \rightarrow -1$.
% Inbetween these two limits 
% $S(a,\lambda_1,\lambda_2)$ has a minumum ($\approx -3.8$ in the plot). 

\subsubsection{Grain Size Distributions}
\label{SecGrainSizeDistributions}

 We concede that a distribution consisting only of a single size of grain may not be the most realistic assumption
for the hypothetical grains trailing KIC 1255b. 
Therefore, we also consider a range of particle sizes -- specifically we consider grains with a specific
particle size distribution denoted by a power-law with slopes, $\nu$, formally defined below.
Power-law size distributions have been used in a diverse range of applications, for example: to model interstellar
dust grains ($\nu$ $\sim$ 3.3 - 3.6; \citealt{Mathis1977}; \citealt{Bierman+Harwit1980}),
% zodiacal dust clouds (INSERTHERE),
large particles
in the comas of comets ($\nu$ $\sim$ 4.7 for particles 1 $\mu m$ to 1 $mm$ in size for the comet 103P/Hartley 2; \citealt{Kelley13}),
and dust (aerosols) in the Earth's lower atmosphere from a world-wide
network of ground-based Sun photometers ($3 \le \nu \le 5$; \citealt{Liou02}; \citealt{Holbenetal1998}).
%We also employ a power-law distribution to consider the putative grains trailing KIC1255b.
%If I assume n of http://arxiv.org/pdf/1304.4204v1.pdf is the same as N(a) in our paper, than nu=-3.7

% Comets give size distribution in terms of dn/da = http://arxiv.org/pdf/1304.4204v1.pdf, or the cumulative number: http://oro.open.ac.uk/22758/1/
We consider a distribution of grains ranging in size from $a = a_{min}$ to $a_{max}$.  Let $N(a)$ be the number density of grains per unit grain radius and per unit area of the column along the path connecting the observer to the star.
The average extinction cross section for the distribution, $\overline{\sigma}_{ext}(a_{min},a_{max},\lambda)$, is given by: 

\begin{equation}
\overline{\sigma}_{ext}(a_{min},a_{max},\lambda) = \frac{ \int_{a_{min}}^{a_{max}} N(a') \, \sigma_{ext}(a',\lambda) \, da' }{ \int_{a_{min}}^{a_{max}} N(a') \, da' }
\label{eqn:SigmaForDistribution01}
\end{equation}
\noindent where the overbar indicates the normalized average over the grain size distribution. 
The definition of \r{A}ngstr\"{o}m exponent is readily adapted to distributions by replacing $\sigma_{ext}(a,\lambda)$ with
 $\overline{\sigma}_{ext}(a_{min},a_{max},\lambda)$ in Eqn.  \ref{eqn:WavelengthDependence01}:

\begin{eqnarray}
%\begin{equation}
\alpha(a_{min},a_{max},\lambda_1,\lambda_2) \equiv \nonumber \\
- \frac{ \log{ [ \overline{\sigma}_{ext}(a_{min},a_{max},\lambda_2) / \overline{\sigma}_{ext}(a_{min},a_{max},\lambda_1) ] } }{ \log{ ( \lambda_2 / \lambda_1 ) } }
\label{eqn:WavelengthDependence02}
%\end{equation}
\end{eqnarray}
\noindent Likewise the first and second columns in Table \ref{TableSensitivityLimits} are readily adapted for size distributions
by replacing $a$ by $\overline{a}$, $a^2$ by $\overline{ a^2}$, and so on. 

\begin{figure}[h]
\includegraphics[width=0.48\textwidth,angle=0]{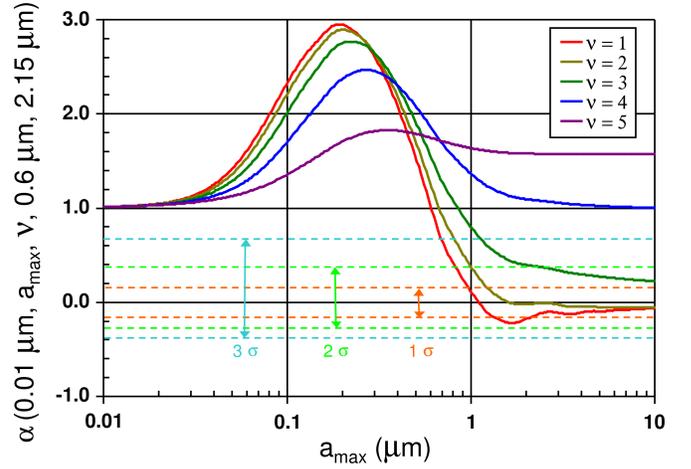}\\
	\caption{
		Plot of extinction sensitivity for a range of particle size distributions with $\nu = 1, \, 2, \, 3, \, 4, \, {\rm and} \, 5$.
		We employ an index of refraction of $n = 1.6$, an imaginary component of the index of refraction, $k = 0.02$, 
		and calculate the particle distribution between a minimum radius of $a_{min} = 0.01 \, {\rm \mu m}$, 
		and a maximum radius of $0.01 \le a_{max} \le 10 \, {\rm \mu m}$.
		The horizontal orange, green, and blue dashed lines show the 1$\sigma$, 2$\sigma$, and 3$\sigma$ limits on $\alpha(a_{min},a_{max},\nu,\lambda_1,\lambda_2)$, respectively,
		from $\S$\ref{SecGrainSizeDistributions}.
		Only smaller power-law slopes, and thus particle size distributions with more large particles, 
		e.g. $\nu = 1, \, 2,$ \& 3, result in a sufficient number of near-micron sized
		grains and thus satisfy the 
		ratio of our CFHT/WIRCam and {\it Kepler} transit depths at the 3$\sigma$ level.
	}
\label{FigPowerLawAngstromPlot}
\end{figure}

We now specialize to the case of a power-law distribution. 
We define the number density of particles of a given size as follows:  $N(a) = {\it c} \, a^{-\nu}$ for $a_{min} \le a \le a_{max}$,
where the particle size distribution parameter, $\nu$, is taken to lie between $1 \le \nu \le 5$, and ${\it c}$ is a normalization
constant\footnote{Incidentally, for aerosols in the Earth's atmosphere, \cite{Junge1963} showed that $\nu$ and $\alpha$ were simply related. An analytic demonstration using a simple approximation is presented by \cite{DeVore2011}.}.
In the following we arbitrarily set the minimum particle size to be $a_{min} = 0.01 \, {\rm \mu m}$,
and set the maximum particle size to lie in an astrophysically plausible range of: $0.01 \, {\rm \mu m} \le a_{max} \le 10 \, {\rm \mu m}$.
We calculate $\alpha(a_{min}=0.01 \, {\rm \mu m}, a_{max}, \nu, 0.6 \, {\rm \mu m},2.15 \, {\rm \mu m})$ for spherical grains comprised of material
having $n = 1.6$ and $k = 0.02$, and plot the results in Figure \ref{FigPowerLawAngstromPlot}.
We note the similarity of this plot to the one for a single-sized particle in 
Fig. \ref{FigSingleParticleAngstromPlot}\footnote{As
$a_{max} \rightarrow 0.01 \, {\rm \mu m}$, $\alpha(a_{min},a_{max},\nu,\lambda_1,\lambda_2) \rightarrow 1$ for all values of $\nu$. As $a_{max}$
increases beyond $\sim 1 \, {\rm \mu m}$, $\alpha(a_{min},a_{max},\nu,\lambda_1,\lambda_2)$ approaches a constant that depends upon $\nu$.
As $\nu$ decreases, the proportion of large particles increases and the constant value approaches $0$.
In between these two limits  $\alpha(a_{min},a_{max},\nu,\lambda_1,\lambda_2)$ has a maximum in the range between roughly $2$ and $3$.}.
%END FOOTNOTE HERE

%
%

%As in Fig. \ref{FigSingleParticleSensitivityPlot}, the dashed red lines in Fig. \ref{fig:PowerLawSensitivityPlot} show the 1 and the dashed green lines
%the 2 standard deviation values below and above the sensitivity value calculated from the measured transit depths in Table \ref{tab:}.
The implication is that there needs to be a sufficient proportion of grains with $a \gtrsim 1 \, {\rm \mu m}$ for the ratio of the extinctions
at the two wavelengths to be so close to $1$. In the case of a power-law grain size distribution this requires both a sufficiently large upper
size cutoff, $a_{max}$, and a sufficiently small power-law slope, $\nu$. Note that in this limit the specific value of the imaginary component of the
index of refraction, $k$, makes little difference. 

At this point we note the important caveat that the wavelength sensitivity as defined in Eqn.~\ref{eqn:WavelengthDependence01}
or \ref{eqn:WavelengthDependence02} implicitly assumes that the dust tail is optically thin. The numerator
in Eqn.~\ref{eqn:SigmaForDistribution01} is the optical thickness, $\tau(\lambda)$, through the dust tail. Using the fact that the 
transmission $T(\lambda) \equiv \exp{[ - \tau(\lambda) ]}$ and that the extinction is $1 - T(\lambda)$, we find that in the optically thin
limit, i.e., where $\tau(\lambda) \ll 1$:
\begin{eqnarray}
%\begin{equation}
1 - T(\lambda) \rightarrow \overline{\sigma}_{ext}(a_{min},a_{max},\lambda) \times  \int_{a_{min}}^{a_{max}} N(a') \, da' \,\,\,\,  \nonumber \\ 
{\rm if \,\, \tau(\lambda) \ll 1~~~}
\label{eqn:OpticallyThinLimit01}
%\end{equation}
\end{eqnarray}
\noindent Since the integral on the right in Eqn.~\ref{eqn:OpticallyThinLimit01} is independent of $\lambda$, extinction is proportional
to $\overline{\sigma}_{ext}(a_{min},a_{max},\lambda)$ in this limit. However, since it is likely that the putative dust tail does not obscure the entire
stellar surface there may be portions of the tail where the dust is optically thick, i.e., $\tau(\lambda) \gtrsim 1$. To the extent that this
is the case, the extinction measurements will tend to give the same value independent of $\lambda$, the assumptions underlying $\alpha(\lambda_1,\lambda_2)$ break down, and
the utility of $\alpha(\lambda_1,\lambda_2)$ for inferring grain size diminishes. 

\subsubsection{The impact of large particles on the supposed forward scattering peak}
The presence of an increase in flux immediately preceding the dip attributable to strong forward
scattering \citep{Rappaport12,Brogi12,Budaj13} is also suggestive of large particles since the strength of forward
scattering is a strong function of grain size \citep[e.g.,][]{DeVoreetal2013}. 
We note that according to the model
of \citet{Budaj13} there may be some tension between our
finding of $a$$\gtrsim \ParticleSizeCare$ ${\rm \mu m}$ particles
and the {\it Kepler} observations, as 0.1 - 1.0 $\mu m$
particles arguably over-predict the amount of forward scattering compared to the
{\it Kepler} observations\footnote{Other possibilities include the suggestion that the cometary tail might 
be composed of different sized particles 
at different distances from the planet, as suggested by \citet{Budaj13},
or that the tail might be composed of particles
with different scattering properties than we would naively expect.}.
However, the forward scattering increase also suggests that the dust
responsible is not optically thick since scattering acts both to inhibit the forward scattered photons as well as cause the 
flux to diverge angularly. If multispectral measurements of the forward scattering peak were obtained with adequate
signal to noise, then they could be used to provide information on grain size through the strong dependence of the angular 
spread of forward scattering on the ratio of the effective grain diameter to the wavelength.

\subsection{KIC 1255b Particle Lifting}

If the largest particles in the tail of KIC 1255b are at
least \ParticleSizeCare \ microns in size, then this raises the
question of how such large particles were lofted from the planet in
the first place. According to the hydrodynamic wind model of
\citet{PerezBeckerChiang13}, the present-day mass of KIC 1255b is
$\lesssim 0.02 M_\oplus$, or less than half the mass of Mercury.  Only
for masses this small are surface gravities weak enough to allow
evaporative winds to blow with mass-loss rates satisfying the
observations. These authors have already shown that spherical grains
having radii of $1$ $\mu$m and bulk densities of $3$ g/cm$^3$
could be dragged by winds outside the Roche lobe of KIC 1255b ---
albeit only marginally if the planet mass is near its upper limit. As
the planet radius shrinks, the sizes of particles that can escape grow
in inverse proportion (see their equation 30 and surrounding
discussion). In sum, as long as the planet is small enough,
micron-sized or larger particles can escape. Therefore, our
observations suggest that the candidate planet KIC 1255b might best be
described as a sub-Mercury --- rather than super-Mercury --- mass
planet.

Note that the particles do not need to be lifted directly from the
planetary surface, since the grains only condense at altitude, where
the wind has an opportunity to cool adiabatically. Indeed the grains
cannot be present at their maximum abundance (relative to gas) at the
base of the wind; if they were, the flow would be so optically thick
that the planetary surface would be shielded from starlight and would
not heat to the temperatures required for an evaporative wind to be
launched\footnote{In principle, the surface could be heated to about 1800 K by
radiation emitted from dusty layers at altitude, but such a surface
temperature would still be too low for the planet to emit a wind of
the required strength to match observations. Heat redistribution by
winds and gravity waves across the day-night terminator would only lower the surface
temperature; see \citet{Budaj12} and \citet{PerezBeckerShowman13}.}.
In the wind solutions of \citet{PerezBeckerChiang13}, the dust
abundance increases by orders of magnitude from the surface to the
Roche lobe (see their Figure 3). Thus the problem of lifting grains
beyond the Roche lobe should only be a problem near the Roche lobe,
and it is there that \citet{PerezBeckerChiang13} compare the drag
force exerted by the wind with the planet's tidally modified
gravity. We suppose it is possible that the problem of lift can be
avoided altogether if grains condense outside the Roche lobe. We
cannot rule this possibility out since the grain condensation profile
was merely parameterized by \citet{PerezBeckerChiang13}, not solved
from first principles.

\subsection{Prospects for follow-up observations}

 The enlightening multi-wavelength observations presented in this paper were greatly facilitated
by the impressive photometric capabilities of the {\it Kepler} spacecraft that allowed us to achieve simultaneous,
accurate optical photometry in addition to our ground and spacebased photometry and imaging.
With the recent malfunction of this spacecraft,
prospects for illuminating follow-up studies of this object are much dimmer than previously, and may require
much more difficult to schedule follow-up from several telescopes simultaneously.

Although the KIC 1255b transit does not appear to be strongly wavelength dependent from 0.6 - 2.15 $\mu m$,
follow-up observations further into the infrared would be expected to show transit depth differences
compared to those obtained simultaneously in the optical, unless
the size of particles in the cometary tail are 
several microns in size. Such observations could be obtained 
with the {\it Spitzer}/IRAC \citep{Fazio04} instrument.
We note that several micron-sized particles 
would be inconsistent with the size of the forward scattering peak observed with
{\it Kepler} \citep{Rappaport12,Budaj13} (unless there is evolution of the grain sizes along the tail).
Low resolution spectroscopy over a wide spectral range, either from the ground or from space, would be useful
to look for both wavelength dependent transit depth changes, and morphological changes in the photometry of the dust tail
as revealed by the forward-scattering peak, and asymmetric egress of the transit.
We note, that the strength of the forward scattering peak 
is also expected to be wavelength dependent, and observations with sufficient precision to look for such minute changes, would
be highly illuminating.

%Spitzer calculations here
%A 13.3 K, K-dwarf should have a 1.4462 mJy (3.6 microns) or 0.8013 mJy (4.5 microns), according to starpet: http://ssc.spitzer.caltech.edu/warmmission/propkit/pet/starpet/
%Using snirac_warm, and assuming a background of MJy/sr = 0.093 (3.6) and MJy/sr = 0.32 (4.5), according to this: http://irsa.ipac.caltech.edu/data/SPITZER/docs/irac/iracinstrumenthandbook/10/
%Results in S/N of 229 (3.6) and 132 (4.5), more than enough precision to detect the transit at these wavelengths...
% You have to remember that this target is relatively bright in the near-infrared (K~13.3), and is thus not that
% faint at the Spitzer wavelengths. I crunched the signal-to-noise calculations using the Spitzer tools and found out
% that we should definitely be able to detect a 1% transit depth. 
% The Ballard et al. observations of Kepler-61 (http://arxiv.org/pdf/1304.6726.pdf) are only a magnitude brighter (K~12.2) and are able to get 0.1% precision
% on timescales of an hour or so. Therefore, I think Spitzer would be useful.
%

% Low-resolution
% spectrographs that achieve high enough precision to  

\section{Conclusions}

 We have presented multiwavelength photometry, high angular-resolution imaging and radial velocities
of the intriguing disintegrating low-mass candidate planet KIC 1255b. We summarize our findings here:

 (i) {\it Comparison of our CFHT/WIRCam 2.15 $\mu m$ to {\it Kepler} 0.6 $\mu m$ transit depths, and the
resulting constraints on particle
sizes in the tail trailing KIC 1255b:} The average ratio of the transit depths
that we observe from the ground with CFHT/WIRCam and space with {\it Kepler}
at our two epochs are: 
\RatioBOTHCFHTKepler \ $\pm$ \RatioBOTHErrorCFHTKepler.
In the disintegrating planet scenario, the
only way to see a lack of extinction from the optical to the near-infrared is if the circumference of the 
particles are at least approximately the wavelength of the observations. Therefore,
if the transits
we observe are due to scattering from single-size particles streaming from the planet in a comet-like tail,
then the particles must be $\sim$\ParticleSizeCare \ microns in radius or 
larger\footnote{
We note this is in some disagreement, and modest agreement, with two efforts
that presented scattering models compared to the {\it Kepler} photometry.
The findings of \citet{Brogi12} modestly disagree with our own, as they 
suggest the particles in the tail trailing KIC 1255b must have a 
a typical grain size of 0.1 $\mu m$ from six quarters of long cadence {\it Kepler} photometry.
The results of \citet{Budaj13} are in modest agreement with our own, as their analysis
of 14 quarters of {\it Kepler} long and short cadence photometry, suggest grain sizes from 0.1 - 1.0 $\mu m$.
}.
Similarly, if the particle size distribution in the tail follows a number density defined by a power-law, then only smaller 
power-law slopes, and thus larger particle sizes result in a sufficient number of near-micron sized
grains to satisfy our observations.

(ii) {\it Comparison of our {\it HST} 1.4 $\mu m$ and {\it Kepler} 0.6 $\mu m$ null detections:} 
Unfortunately we were unable to detect the transit of KIC 1255b in either 
our simultaneous {\it HST} and {\it Kepler} photometry, due to the fact that the transits of KIC 1255b had largely disappeared
in the {\it Kepler} photometry for approximately $\sim$5 days before and after our observed transit. We are therefore
able to conclude little from these observations, other than there is no evidence for strongly different transit depths
at these wavelengths.

(iii) {\it Particle lifting from KIC 1255b:} \citet{PerezBeckerChiang13} have already demonstrated that lifting particles
nearly a micron in size is possible from KIC 1255b. As lifting such large particles becomes much more difficult as one
increases the mass of the candidate planet, we note our $\gtrsim$\ParticleSizeCare \ micron limit on single-sized particles
in the tail trailing KIC 1255b favours a sub-Mercury, rather than super-Mercury, mass for KIC 1255b.

(iv) {\it Constraints on false-positives from our high angular-resolution imaging, RVs and photometry:} Our {\it HST}
($\sim$0.53 $\mu m$ and $\sim$0.77 $\mu m$) high angular-resolution imaging allows us to 
rule-out
background and foreground candidates at angular separations greater than 0.2\arcsec \ that could be responsible
for the transit we associate with a planet transiting KIC 1255b. The associated limit 
from our 
groundbased Keck/NIRC2 Adaptive Optics observations in K'-band ($\sim$2.12 $\mu m$) is 
for separations greater than 1.4\arcsec. Our radial velocity observations
allow us to rule out low-mass stellar companions
($\gtrsim$ 0.2 M$_\odot$) for periods less than $\lesssim$$10$ years,
and 13 Jupiter-mass companions for periods less than $\lesssim$$40$ days.
Furthermore, the similar transit depths we observe in the near-infrared with CFHT/WIRCam and in the optical
with {\it Kepler} also allow us to rule out background/foreground candidates, or higher-order multiples with 
significantly different spectral types, as this would result in a colour-dependent transit depth from the optical
to the near-infrared.
Although prior to these observations we knew of no viable false-positive scenario
that could reproduce the unique photometry we observed with {\it Kepler} 
(e.g. the forward scattering bump before transit, the sharp ingress and gradual egress transit profile,
the sharply varying transit depths), we note that we have now greatly reduced the parameter space for viable false positive scenarios.
We conclude that the disintegrating low-mass planet scenario is the simplest explanation for our multiwavelength
photometry, RVs and high angular-resolution imaging suggested to date. 

(v) {\it Limit on the mass of the candidate planet KIC 1255b:} Our KECK/HIRES RVs of
KIC 1255b allow us to place an upper-limit on the minimum mass of the candidate
planet that confirms it is firmly in the planetary regime; this limit
is $m \, \sin i$ $\lesssim$ 1.2 M$_J$ with 3$\sigma$ confidence, assuming a circular orbit. %PLANETMASSLIMIT

\acknowledgements

B.C.'s work was performed under a contract with the California Institute of
Technology funded by NASA through the Sagan Fellowship Program.
The Natural Sciences and Engineering Research Council of Canada supports the research of B.C.
Support for program \#GO-12987
was provided by NASA through a grant from the Space Telescope Science Institute, which is operated by the
Association of Universities for Research in Astronomy, Inc., under NASA contract NAS 5-26555
This work was based on observations at the W. M. Keck Observatory granted
by the University of Hawaii and the University of California. We thank
the observers who contributed to the measurements reported here and
acknowledge the efforts of the Keck Observatory staff. We extend special
thanks to those of Hawaiian ancestry on whose sacred mountain of Mauna
Kea we are privileged to be guests. 
% The authors wish to recognize and acknowledge the very significant cultural role and reverence that the summit of Mauna Kea has always had within
% the indigenous Hawaiian community.  We are most fortunate to have the opportunity to conduct observations from this mountain.
The authors thank the referee, Jan Budaj, for helpful comments that have improved this manuscript,
and for providing us with very convenient tables of
indices of refraction for various minerals.
The authors especially appreciate the hard-work and diligence of the CFHT staff
for both scheduling the challenging CFHT observations described here and ensuring these ``Staring Mode'' observations were successful.
The authors thank Geoff Marcy for contributing to and assisting with
the Keck/HIRES RV observations of KIC 1255b that we discuss in this work.
We also thank Ray Jayawardhana, David Lafreniere, 
Magali Deleuil, and Claire Moutou for contributing to the CFHT observing proposal, and Josh Winn
for contributing to the {\it HST} observing proposal,
on which this work is partially based.

\end{document}